\documentclass[aps,prx,reprint,nofootinbib,preprintnumbers,floatfix,unsortedaddress,superscriptaddress,longbibliography]{revtex4-2}

\input{header-1-packages}

\newcommand{\ungroup}[1]{#1}
\newcommand{\withbreak}[1]{\expandafter\ungroup#1}

\def\dounbracket[#1]{#1}

\newcommand{\SU}{{\mathrm{SU}}}

\newcommand{\ubar}{{\bar{u}}}
\newcommand{\dbar}{{\bar{d}}}

\renewcommand\vec\mathbf

\newcommand{\cD}{{\mathcal{D}}}
\newcommand{\sD}{{\mathcal{D}_s}}

\newcommand\redsout{\bgroup\markoverwith{\textcolor{red}{\rule[0.5ex]{2pt}{1.4pt}}}\ULon}

\newcommand\dootimesall[2]{\ifx0#1\else\mathbf{#1}\ifx0#2\else\def\mytmp{\otimes\dootimesall{#2}}\expandafter\expandafter\expandafter\mytmp\fi\fi}

\usepackage{relsize}

\usepackage[acronyms, nohypertypes={acronym}, nopostdot, style=super, nonumberlist, toc]{glossaries}
\setacronymstyle{long-sc-short}
\glsenableentrycount

\newcommand\ac[1]{\gls{#1}}

\newacronym{PNA}{pna}{particle-number-algorithm}
\newacronym{SFA}{sfa}{spin-flip-algorithm}

\newacronym{WF}{wf}{Wilson-Fisher}
\newacronym{AF}{af}{asymptotically free}

\newacronym{RG}{rg}{renormalization group}

\newacronym{QIS}{qis}{Quantum Information Science}
\newacronym{PPT}{ppt}{positive-semidefinite partial transpose}
\newacronym{KS}{ks}{Kogut-Susskind}
\newacronym{NPT}{npt}{negative partial transpose}
\newacronym{SMG}{smg}{symmetric mass generation}
\newacronym{SMF}{smf}{symmetric massive fermion}
\newacronym{MF}{mf}{massless fermion}
\newacronym{SSB}{ssb}{spontaneous symmetry breaking}

\newacronym{AS}{as}{Anti-Symmetric}

\newacronym[longplural={conformal field theories}]{CFT}{cft}{conformal field theory}
\newacronym[longplural={lattice field theories}]{LFT}{lft}{lattice field theory}
\newacronym[longplural={effective field theories}]{EFT}{eft}{effective field theory}
\newacronym[longplural={quantum field theories}]{QFT}{qft}{quantum field theory}
\newacronym[longplural={lattice gauge theories}]{LGT}{lgt}{lattice gauge theory}
\newacronym[longplural={monomer-dimer tensor-networks}]{MDTN}{mdtn}{monomer-dimer tensor-network}

\newacronym[]{DMRG}{dmrg}{Density Matrix Renormalization Group}
\newacronym[]{TFIM}{tfim}{Transverse Field Ising Model}

\newacronym[]{LOCC}{locc}{Local Operations and Classical Communicaton}
\newacronym[]{OBC}{obc}{open boundary conditions}

\newacronym{MPS}{mps}{matrix product states}

\newacronym{JLP}{jlp}{Jordan-Lee-Preskill}

\newacronym{BBN}{bbn}{big bang nucleosynthesis}

\newacronym{LEC}{lec}{low-energy constant}

\newacronym{QCD}{qcd}{quantum chromodynamics}
\newacronym{MC}{mc}{Monte Carlo}

\newacronym{IR}{ir}{infrared}
\newacronym{UV}{uv}{ultraviolet}

\newacronym{QED}{qed}{quantum electrodynamics}
\newacronym{SNR}{snr}{signal-to-noise ratio}

\newacronym{NLSM}{nlsm}{nonlinear sigma model}

\newacronym{CL}{cl}{Complex Langevin}

\newacronym{CSA}{csa}{Cartan subalgebra}

\newacronym{AFQMC}{afqmc}{auxiliary field quantum Monte Carlo}
\newacronym{iHMC}{ihmc}{imaginary-mass Hybrid Monte Carlo}

\newacronym{MCMC}{mcmc}{Markov Chain Monte Carlo}

\newacronym{QI}{qi}{quantum information}

\newacronym{irrep}{{\rm irrep}}{irreducible representation}

\newacronym{ASQR}{asqr}{antisymmetric qubit regularization}

\begin{document}

\title{Phase diagram of a lattice fermion model with symmetric mass generation}
\author{Sandip Maiti\,\orcidlink{0000-0002-5248-5316}}
\email{sandipmaiti73@gmail.com}
\affiliation{Key Laboratory of Quark and Lepton Physics (MOE) and Institute of Particle Physics, Central China Normal University, Wuhan 430079, China}
\affiliation{Saha Institute of Nuclear Physics, HBNI, 1/AF Bidhannagar, Kolkata 700064, India}
\author{Debasish Banerjee\,\orcidlink{0000-0003-0244-4337}}
\email{D.Banerjee@soton.ac.uk}
\affiliation{Saha Institute of Nuclear Physics, HBNI, 1/AF Bidhannagar, Kolkata 700064, India}
\affiliation{School of Physics and Astronomy, University of Southampton, University Road, SO17 1BJ, UK}
\author{Shailesh Chandrasekharan\,\orcidlink{0000-0002-3711-4998}}
\email{sch27@duke.edu}
\affiliation{Department of Physics, Box 90305, Duke University, Durham, North Carolina 27708, USA}
\author{Marina K.~Marinkovic\,\orcidlink{0000-0002-9883-7866}}
\email{marinama@ethz.ch}
\affiliation{Institut f\"ur Theoretische Physik, Wolfgang-Pauli-Stra{\ss}e 27, ETH Z\"urich, 8093 Z\"urich, Switzerland}

\date{\today}

\begin{abstract}
We study the phase structure of a model containing two flavors of massless staggered fermions interacting through two independent four-fermion couplings, $U_I$ and $U_B$, formulated on a three-dimensional Euclidean space–time lattice. At $U_B = 0$, this model is known to exhibit a direct second-order quantum phase transition between a \ac{MF} phase and a phase in which fermions acquire masses through the mechanism commonly referred to as \ac{SMG}. We demonstrate that introducing a small nonzero value of $U_B$ qualitatively alters this structure: the single exotic transition at $U_B = 0$ splits into two distinct, conventional transitions, separated by an intermediate phase in which fermion masses arise through the standard mechanism of \ac{SSB}. The first of these is a Gross–Neveu transition separating the \ac{MF} phase from the \ac{SSB}-induced massive phase, while the second is a three-dimensional XY transition between the \ac{SSB} phase and the \ac{SMG} phase. Using the fermion-bag Monte Carlo method, we verify that the critical exponents associated with both transitions are consistent with the literature, thereby yielding a quantitative characterization of the resulting phase structure of the model.
\end{abstract}

\maketitle

\section{Introduction}
\label{sec-1}

The origin of fermion masses is a fundamental question in particle physics. In quantum field theories, such masses typically arise from local fermion bilinear terms in the action. When symmetries forbid these terms, fermion mass generation usually proceeds through spontaneous symmetry breaking (\ac{SSB}), such that fermion bilinear terms emerge dynamically as condensates \cite{Gross:1974jv}. A similar mechanism is employed in the Standard Model, where the Higgs field develops a vacuum expectation value and thereby enables fermion masses to arise through Yukawa couplings \cite{Logan:2014jla,Pich:2012sx}. In Quantum Chromodynamics, \ac{SSB} of chiral symmetry is driven by strong gauge--field fluctuations, which induce fermion bilinear condensates; together with confinement, this mechanism accounts for the large masses of hadrons such as protons and neutrons.

However, recent studies of lattice models have demonstrated that fermion masses can also emerge dynamically without \ac{SSB} \cite{Slagle:2014vma,PhysRevX.8.011026,PhysRevX.11.011063}. In such cases, none of the lattice symmetries present in the massless phase are spontaneously broken in the massive phase, and the masses instead arise purely from strong interactions. This unconventional mechanism of fermion mass generation is now commonly referred to as \ac{SMG} \cite{PhysRevX.8.011026,Tong:2021phe,Wang:2022ucy}. Although \ac{SMG}-like phases were discovered long ago \cite{Lee:1989mi,Bock:1990tv}, they were largely overlooked for many years, having been assumed to be lattice artifacts without relevance to continuum quantum field theories. The recent discoveries of a direct second--order transition between the \ac{MF} phase and the \ac{SMG} phase \cite{Ayyar:2014eua,Ayyar:2015lrd,Ayyar:2016lxq,Catterall:2015zua,He:2016sbs, Maiti:2025ley} have reignited interest in the \ac{SMG} phase itself \cite{Butt:2018nkn,Butt:2021koj,Butt:2024kxi,Hasenfratz:2025lti}, as well as renewed efforts to formulate chiral gauge theories on the lattice using \ac{SMG} to gap out mirror fermions \cite{Eichten:1985ft,Wang:2018ugf,Xu:2021ztz,Catterall:2022jky,Zeng:2022grc,Golterman:2023zqf,Golterman:2025boq}.

The presence of a direct second--order transition in a lattice model between the \ac{MF} phase and the \ac{SMG} phase suggests the existence of a new RG fixed point that governs the long--distance physics at criticality. Such a fixed point, featuring two relevant directions, was conjectured long ago in a class of lattice models \cite{Hasenfratz:1988vc}. The most convincing evidence for its existence now comes from studies in three Euclidean space--time dimensions on a cubic lattice \cite{Ayyar:2014eua,Ayyar:2015lrd}. Somewhat surprisingly, however, a single four-fermion interaction parameter---which we label $U_I$---proved sufficient to tune the model to criticality and reveal the transition \cite{Ayyar:2014eua,Ayyar:2015lrd}.

Motivated by the possibility that an underlying symmetry may reduce the number of relevant directions, in this work we extend the lattice model studied previously by introducing a new coupling $U_B$ that breaks some of the lattice symmetries. To explore the effects of $U_B$ on the phase transition, we perform large-scale Monte Carlo calculations, in the absence of explicit fermion mass terms in the lattice action. A key feature of our construction is that the model admits a formulation within the fermion-bag approach \cite{Chandrasekharan:2009wc}, which enables efficient simulations in this regime.

Our results show that the presence of $U_B$ splits the direct transition between the \ac{MF} phase and the \ac{SMG} phase into two distinct transitions by generating an intermediate, more conventional massive-fermion phase characterized by \ac{SSB}. The transition between the \ac{MF} phase and the \ac{SSB} phase belongs to a conventional Gross--Neveu universality class, while the transition between the \ac{SSB} phase and the \ac{SMG} phase belongs to a bosonic universality class in which fermions are massive and decouple. 
The symmetries of our model and the associated symmetry-breaking pattern suggest that this transition lies in the three-dimensional XY universality class. In our work, we are able to study the universality classes of both transitions and find critical behavior consistent with these expectations.

Our paper is organized as follows. In \cref{sec-2}, we introduce our lattice model and discuss its relevant lattice symmetries. In \cref{sec-3}, we show how we can formulate the fermionic path integral of our model using the fermion bag approach. We introduce the idea of instanton-dimer configurations and their weights. We then construct several Monte Carlo methods to update these instanton-dimer configurations in \cref{sec-4}. In \cref{sec-5}, we present our results. We first discuss a detailed study of the Gross-Neveu transition followed by a discussion of the XY-transision. We then show a broader study of the model for various couplings and lattice sizes to obtain a qualitative understanding of the phase diagram of our model. Each of these discussions are presented as separate subsections of \cref{sec-5}. We finally present our conclusions in \cref{sec-6}.

\section{Lattice Model}
\label{sec-2}

Our lattice model is a $(2+1)$-dimensional Euclidean space--time system consisting of two flavors of massless staggered fermions, denoted $u$ and $d$, interacting via two independent four-fermion couplings $U_I$ and $U_B$. The action is constructed with four Grassmann-valued staggered fermion fields at each site $i$, namely $\ubar_i$, $u_i$, $\dbar_i$, and $d_i$, and is given by
\begin{widetext}
\begin{align}
S
= \sum_{\langle ij\rangle} \frac{\eta_{ij}}{2}\,\bigl(\ubar_i u_j - \ubar_j u_i + \dbar_i d_j - \dbar_j d_i\bigr)
\;-\; U_I \sum_{i} \bigl( \ubar_i u_i \,\dbar_i d_i \bigr)
\;-\; U_B \sum_{\langle ij\rangle} 
\bigl( \ubar_i u_i\, \ubar_j u_j 
      + \dbar_i d_i\, \dbar_j d_j \bigr).
\label{eq:model}
\end{align}
\end{widetext}
Here $i$ and $j$ label sites on a cubic lattice with coordinates $(t,x,y)$. Each side of the cube has length $L$ in lattice units, so that $t,x,y = 0,1,\ldots,L-1$, and the total lattice volume contains $L^3$ sites. The notation $\langle ij\rangle$ denotes the nearest-neighbor lattice sites. The phase factors $\eta_{ij}$ implement a $\pi$ flux through each plaquette of the cubic lattice. A convenient choice is
$\eta_{ij} = 1$ if $j$ has coordinates $(t+1,x,y)$,
$\eta_{ij} = (-1)^t$ if $j$ has coordinates $(t,x+1,y)$,
and $\eta_{ij} = (-1)^{t+x}$ if $j$ has coordinates $(t,x,y+1)$.
Finally, lattice sites are classified as even or odd depending on whether $t+x+y$ is even or odd, respectively. In this work, anti-periodic boundary conditions are imposed along each direction.

We can view the action in \cref{eq:model} as consisting of three distinct terms, each with its own symmetry properties. The first term describes the free hopping of fermions and is invariant under an internal $SU(4)\times U(1)$ transformation given by
\begin{align}
\begin{pmatrix}
u_i \\
\ubar_i \\
d_i \\
\dbar_i
\end{pmatrix} 
\;\to\; V\, e^{i\theta}
\begin{pmatrix}
u_i \\
\ubar_i \\
d_i \\
\dbar_i
\end{pmatrix},
\qquad
\begin{pmatrix}
\ubar_j \\
u_j \\
\dbar_j \\
d_j 
\end{pmatrix} 
\;\to\; V^* e^{-i\theta}
\begin{pmatrix}
\ubar_j \\
u_j \\
\dbar_j \\
d_j 
\end{pmatrix},
\label{eq:symm-1}
\end{align}
where site $i$ is assumed to be even and $j$ odd. Here \(V\) denotes an element of the \(SU(4)\) group, while \(e^{i\theta}\) represents a \(U(1)\) phase factor often interpreted as an axial symmetry of staggered fermions.

The second term proportional to~$U_I$, is invariant under the \(SU(4)\) transformation in \cref{eq:symm-1}, but it explicitly breaks the axial \(U(1)\) symmetry. For this reason, this interaction can be viewed as a ’t~Hooft vertex, introducing instanton-like effects into the fermionic path integral \cite{Chandrasekharan:2016enx}; this interpretation also motivates the notation \(U_I\). The third term  proportional to~$U_B$, represents back--and--forth hopping of quarks along a single bond, motivating the label \(U_B\). This interaction has also been interpreted as a Thirring interaction for each flavor in previous studies \cite{PhysRevLett.108.140404}. The term is invariant under two independent \(SU(2)\times U(1)\) transformations, one acting on each flavor, both of which are embedded as subgroups of the full \(SU(4)\times U(1)\) symmetry of \cref{eq:symm-1}. The transformation acting on the $u$ quarks takes the form
\begin{align}
\begin{pmatrix}
u_i \\
\ubar_i
\end{pmatrix}
\;\to\; V_u\, e^{i\theta_u}
\begin{pmatrix}
u_i \\
\ubar_i
\end{pmatrix},
\qquad
\begin{pmatrix}
\ubar_j \\
u_j
\end{pmatrix}
\;\to\; V_u^* e^{-i\theta_u}
\begin{pmatrix}
\ubar_j \\
u_j
\end{pmatrix},
\label{eq:symm-2}
\end{align}
where \(V_u \in SU(2)\), and as before $i$ is an even site and $j$ is an odd site. The second \(SU(2)\times U(1)\) symmetry acting on the $d$ quarks is identical to \cref{eq:symm-2}, but with $V_u \to V_d$ and $e^{i\theta_u} \to e^{i\theta_d}$.

It is straightforward to verify that the \(U(1)\) symmetry of \cref{eq:symm-1} is recovered by setting 
\(\theta = \theta_u = \theta_d\).
The other independent \(U(1)\) symmetry of the third term is obtained by choosing 
\(\chi = \theta_u = -\theta_d\); this corresponds to an additional chiral symmetry of two-flavor staggered fermions, which we denote by \(U_\chi(1)\). This symmetry is a subgroup of the \(SU(4)\) symmetry appearing in \cref{eq:symm-1}. 
When only \(U_I\) is nonzero, the symmetry of the model in \cref{eq:model} is the full \(SU(4)\).
When both \(U_I\) and \(U_B\) are nonzero, this \(SU(4)\) symmetry is reduced to 
\(SU(2)\times SU(2)\times U_\chi(1)\times Z_2\), where the latter $Z_2$ symmetry exchanges $u$ and $d$ fermions.

Fermion bilinear condensates of the form $\langle \ubar u\rangle$ and $\langle \dbar d\rangle$ spontaneously break the $U_\chi(1)$ symmetry. Such condensates can be detected by measuring susceptibilities involving the fermion bilinear fields $\ubar_i u_i$ and $\dbar_i d_i$. We will define three susceptibilities of this type:
\begin{align}
\chi_{ud} \;=\; \frac{1}{2L^3}\sum_{i,j} \big\langle \ubar_i u_i \;\dbar_j d_j \big\rangle,
\label{eq:obs-chiud} \\
\chi_{uu} \;=\; \frac{1}{2L^3}\sum_{i,j} \big\langle \ubar_i u_i \;\ubar_j u_j \big\rangle,
\label{eq:obs-chiuu} \\
\chi_{dd} \;=\; \frac{1}{2L^3}\sum_{i,j} \big\langle \dbar_i d_i \;\dbar_j d_j \big\rangle,
\label{eq:obs-chidd}
\end{align}
where $\langle \cdot \rangle$ denotes the usual path-integral expectation value using the action in \cref{eq:model}.  
By symmetry of the action, it is easy to verify that $\chi_{uu} = \chi_{dd}$. We also find empirically that $\chi_{ud} \approx \chi_{uu}$ on large lattices, except when $U_I$ is small. One way to understand this equality is to note that when $i$ and $j$ are well separated, the cluster decomposition property of correlation functions implies
\begin{align}
\langle \bar{u}_i u_i \bar{u}_j u_j\rangle &\approx \langle \bar{u}_i u_i\rangle \langle\bar{u}_j u_j\rangle,
\nonumber\
\langle \bar{u}_i u_i \bar{d}_j d_j\rangle &\approx \langle \bar{u}_i u_i\rangle \langle\bar{d}_j d_j\rangle.
\end{align}
Furthermore, the $Z_2$ symmetry that exchanges the $u$ and $d$ fermions implies
\begin{align}
\langle\bar{u}_j u_j\rangle = \langle\bar{d}_j d_j\rangle.
\end{align}
Combining these observations, we obtain
\begin{align}
\langle \bar{u}_i u_i \bar{u}_j u_j\rangle \approx \langle \bar{u}_i u_i \bar{d}_j d_j\rangle,
\end{align}
whenever $i$ and $j$ are sufficiently far apart. Consequently, we expect $\chi_{uu}$ and $\chi_{ud}$ to be approximately equal when the dominant contributions to the sums in \cref{eq:obs-chiuu} and \cref{eq:obs-chiud} come from widely separated pairs $(i,j)$. This regime is expected to be realized when $U_I$ is sufficiently large.

In a phase where the fermion bilinear condensates are nonzero, we expect these susceptibilities to diverge as $\chi \sim L^3$, while in both the \ac{MF} phase and the \ac{SMG} phase we expect $\chi \sim \mathrm{const.}$  
At a critical point, in contrast, we expect the finite-size scaling behavior $\chi \sim L^{2-\eta}$.

\section{Fermion Bag Approach}
\label{sec-3}

In this work we extend the ideas of the fermion bag approach \cite{Chandrasekharan:2009wc} to construct a Monte Carlo method capable of computing the susceptibilities defined in \cref{eq:obs-chiud,eq:obs-chiuu,eq:obs-chidd}. The path--integral expectation value of any Grassmann-valued observable \(O\) is defined as
\begin{align}
\langle O \rangle
\;=\;
\frac{1}{Z}
\int [\cD\dbar\, \cD d\, \cD\ubar\, \cD u]\;
O\;
e^{-S},
\label{eq:obs-def}
\end{align}
where the partition function is
\begin{align}
Z
\;=\;
\int [\cD\dbar\, \cD d\, \cD\ubar\, \cD u]\;
e^{-S},
\label{eq:pf-def}
\end{align}
and the action \(S\) is given in \cref{eq:model}.

\begin{figure}[h]
\centering
\includegraphics[width=0.43\textwidth]{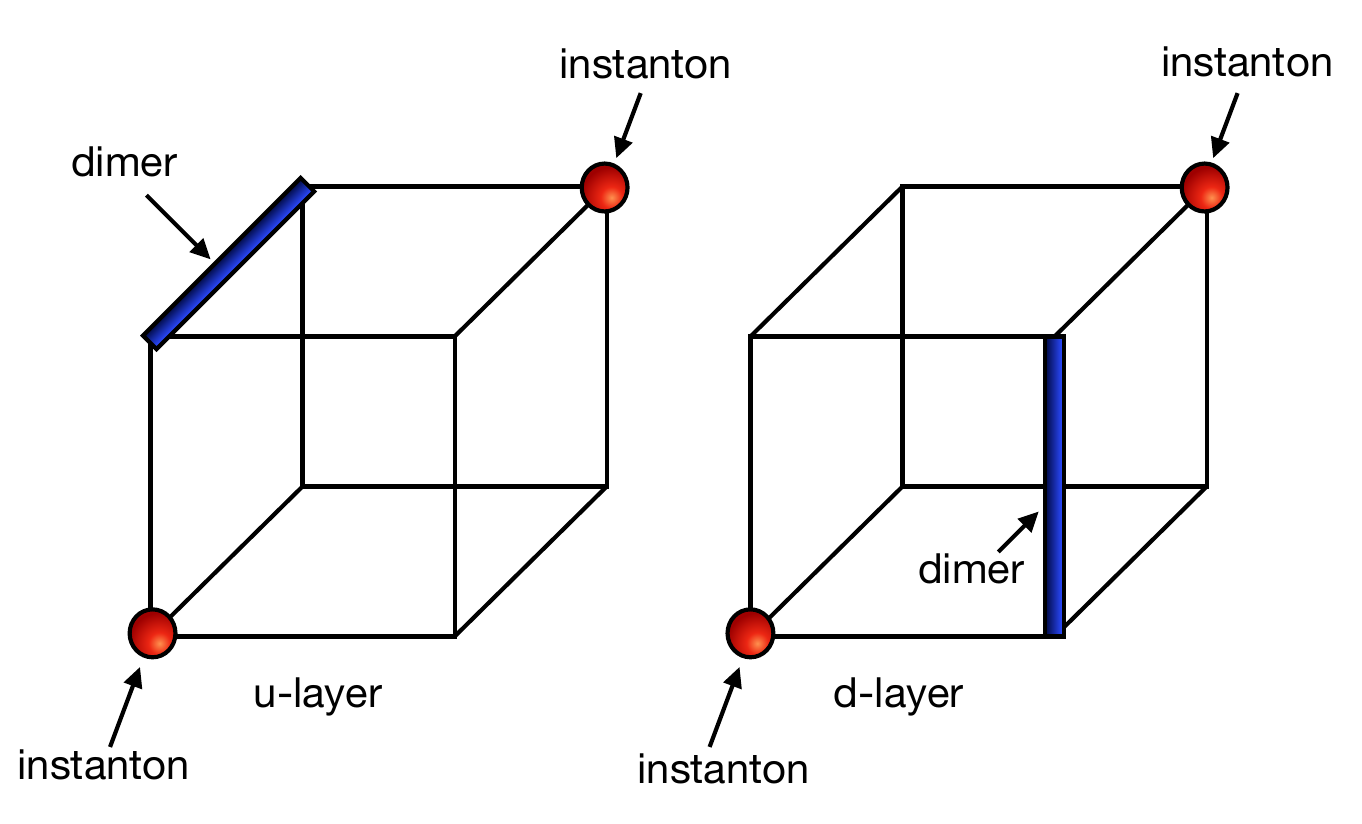}
\caption{An illustration of an instanton--dimer configuration $[n_i,n_b^u,n_b^d]$ on an $L=2$ lattice. Instantons are shown as solid circles, and dimers are shown as thick bonds. The configuration on the left depicts the instantons and dimers on the $u$ layer, while the configuration on the right shows the corresponding structure on the $d$ layer. Instantons occupy the same lattice sites on both layers, whereas the dimer configurations may differ. Sites containing neither instantons nor dimers are also present.\label{fig:fbagL2}}
\end{figure}

The key idea of the fermion bag approach is to reorganize the expression in \cref{eq:obs-def} as a sum over contributions from configurations of \emph{fermion bags}. Each fermion bag is a set of lattice sites within which fermions are allowed to hop. There are two complementary viewpoints for constructing the fermion bag representation: the \emph{weak--coupling} viewpoint and the \emph{strong--coupling} viewpoint \cite{Chandrasekharan:2013rpa}. In the weak--coupling viewpoint, fermion bags form nonlocal sets of sites and fermion propagation occurs through free--fermion propagators; this formulation is equivalent to the diagrammatic Monte Carlo method \cite{Kozik:2010zz}. In the strong--coupling viewpoint, by contrast, fermions hop only locally within connected regions of the lattice. In this work we adopt the weak--coupling viewpoint.

To formulate the fermion bag approach we reorganize the expression in \cref{eq:obs-def}, by first rewriting the kinetic term as
\begin{align}
\sum_{\langle ij\rangle} \frac{\eta_{ij}}{2}
\bigl(\ubar_i u_j - \ubar_j u_i
& + \dbar_i d_j - \dbar_j d_i
\bigr) \notag \\
& \;=\;
\ubar_i M_{ij} u_j + \dbar_i M_{ij} d_j ,
\end{align}
where summation over lattice sites \(i,j\) is implicit on the right hand side. We have introduced the fermion matrix \(M\), which can be identified as an \(L^3\times L^3\) antisymmetric matrix describing a single flavor of free staggered fermions. Note that \(M\) has nonzero entries only between sites of opposite parity and is anti-symmetric.

Using this notation, the partition function may be written as
\begin{align}
Z
&=
\int [\cD\dbar\, \cD d\, \cD\ubar\, \cD u]\;
e^{-\bigl(\ubar_i M_{ij} u_j + \dbar_i M_{ij} d_j\bigr)}
\notag \\
&\times
\prod_i
e^{U_I \ubar_i u_i\, \dbar_i d_i}
\prod_{b\equiv\langle ij\rangle}
e^{U_B (\ubar_i u_i\, \ubar_j u_j + \dbar_i d_i\, \dbar_j d_j)} ,
\label{eq:pf-1}
\end{align}
where we have substituted the action from \cref{eq:model} into \cref{eq:pf-def}. For convenience, we denote each nearest--neighbor bond \(\langle ij\rangle\) by \(b\).

We now expand the interaction terms using
\begin{align}
e^{U_I \ubar_i u_i\, \dbar_i d_i}
&=
\sum_{n_i=0,1}
\bigl(U_I \ubar_i u_i\, \dbar_i d_i\bigr)^{n_i}, \\
e^{U_B \ubar_i u_i\, \ubar_j u_j}
&=
\sum_{n_b^u=0,1}
\bigl(U_B \ubar_i u_i\, \ubar_j u_j\bigr)^{n_b^u}, \\
e^{U_B \dbar_i d_i\, \dbar_j d_j}
&=
\sum_{n_b^d=0,1}
\bigl(U_B \dbar_i d_i\, \dbar_j d_j\bigr)^{n_b^d},
\end{align}
where \(n_i\) denotes the presence of an instanton at site \(i\), and
\(n_b^u\), \(n_b^d\) denote the presence of a \(u\)-type or \(d\)-type dimer on bond \(b\). The set of all $n_i$, $n_b^u$ and $n_b^d$ define a instanton-dimer configuration. An illustration of such a configuration on a $L=2$ lattice is shown in \cref{fig:fbagL2}.

Using these expansions, \cref{eq:pf-1} can be reorganized as a sum over instanton-dimer configurations
\([n_i,n_b^u,n_b^d]\):
\begin{widetext}
\begin{align}
Z
&=
\sum_{[n_i,n_b^u,n_b^d]}
\int [\cD\dbar\, \cD d\, \cD\ubar\, \cD u]\;
e^{-\bigl(\ubar_i M_{ij} u_j + \dbar_i M_{ij} d_j\bigr)}
\prod_i (U_I \ubar_i u_i\, \dbar_i d_i)^{n_i}
\prod_b
(U_B \ubar_i u_i\, \ubar_j u_j)^{n_b^u}
(U_B \dbar_i d_i\, \dbar_j d_j)^{n_b^d}
\notag \\
&=
\sum_{[n_i,n_b^u,n_b^d]}
U_I^{n_I} U_B^{n_B}\quad \Big\{
\int [\cD\ubar\, \cD u]\;
e^{-\ubar_i M_{ij} u_j}
\prod_i (\ubar_i u_i)^{n_i}
\prod_b (\ubar_i u_i\, \ubar_j u_j)^{n_b^u}\Big\}
\notag \\
& \hspace{2in} \times \quad \Big\{
\int [\cD\dbar\, \cD d]\;
e^{-\dbar_i M_{ij} d_j}
\prod_i (\dbar_i d_i)^{n_i}
\prod_b (\dbar_i d_i\, \dbar_j d_j)^{n_b^d}\Big\},
\label{eq:pf-2}
\end{align}
\end{widetext}
where we have separated the Grassmann integrals over the \(u\) and \(d\) fields, and defined
\begin{align}
n_I = \sum_i n_i,
\qquad
n_B = \sum_b (n_b^u + n_b^d).
\end{align}

Each Grassmann integral can be evaluated using Wick’s theorem. In both cases, the result is a determinant of a matrix \(G\) whose elements are entries of \(M^{-1}\) connecting the sites where either \(n_i = 1\) or where the site is attached to a bond \(b\) with \(n_b = 1\). Further details may be found in Ref.~\cite{Chandrasekharan:2013rpa}. Using this result, we can show that the integral over the $u$ fields is given by
\begin{align}
\int [\cD\ubar\, \cD u]\;
e^{-\ubar_i M_{ij} u_j}
& \prod_i (\ubar_i u_i)^{n_i}
\prod_b (\ubar_i u_i\, \ubar_j u_j)^{n_b^u}
\notag \\
&= \det(M) \det(G_u),
\end{align}
where \(G_u\) is a square matrix of dimension \(n_I + 2 n_b^u\). Similarly, the integral over the $d$ fields yields \(\det(M) \det(G_d)\), where \(G_d\) has dimension \(n_I + 2 n_b^d\). The final expression for the partition function is therefore 
\begin{align}
Z
\;=\; (\det M)^2
\sum_{[n_i,n_b^u,n_b^d]}
U_I^{n_I} U_B^{n_B}
\det(G_u)\, \det(G_d).
\label{eq:pf-fb}
\end{align}
Since both matrices are antisymmetric, their determinants are nonnegative. Consequently, the partition function is expressed as a sum of positive terms, which can be efficiently sampled using a Monte Carlo method.

We can repeat the above derivation for the observable defined in \cref{eq:obs-def}, obtaining
\begin{align}
\langle O \rangle\, Z
\;=\; (\det M)^2
\sum_{[n_i,n_b^u,n_b^d;n_O]}
U_I^{n_I} U_B^{n_B}
\det(\tilde{G}_u)\, \det(\tilde{G}_d),
\label{eq:obs-fb}
\end{align}
where \(\tilde{G}_u\) and \(\tilde{G}_d\) are modified matrices that incorporate the Grassmann fields appearing in the observable \(O\). These fields also modify the instanton-dimer configurations to a new class of configurations which we label as $[n_i,n_b^u,n_b^d;n_O]$.

For example, in the computation of \(\chi_{ud}\) using \cref{eq:obs-chiud}, we consider the observable
\(O = \ubar_k u_k\, \dbar_\ell d_\ell\).
We refer to factors such as \(\ubar_k u_k\) as a \emph{$u$-type monomer} at site \(k\), and similarly \(\dbar_\ell d_\ell\) as a \emph{$d$-type monomer} at site \(\ell\). In this case, the matrices \(\tilde{G}_u\) and \(\tilde{G}_d\) are constructed from an extended instanton--dimer--monomer configuration, denoted by
\([n_i, n_b^u, n_b^d; k, \ell]\),
in which the $u$ layer contains an additional monomer at site \(k\) and the $d$ layer contains an additional monomer at site \(\ell\). Specifically, the $u$-field integral takes the form
\begin{align}
\int [\cD\ubar\, \cD u]\;
e^{-\ubar_i M_{ij} u_j}
& \prod_i (\ubar_i u_i)^{n_i}
\prod_b (\ubar_i u_i\, \ubar_j u_j)^{n_b^u}
\;\ubar_k u_k
\notag \\
&= \det(M) \det(\tilde{G}_u),
\end{align}
and similarly \(\tilde{G}_d\) is defined by inserting \(\dbar_\ell d_\ell\) into the corresponding $d$-field integral. The full computation of \(\chi_{ud}\) therefore involves summing over configurations with one $u$-type monomer and one $d$-type monomer inserted at all possible lattice sites, followed by division by the factor \(L^3\) as specified in \cref{eq:obs-chiud}. Similarly, \(\chi_{uu}\) involves summing over configurations with two $u$-type monomers inserted at all possible lattice sites and dividing by the factor \(2L^3\), in accordance with \cref{eq:obs-chiuu}.

Using \cref{eq:obs-fb,eq:pf-fb}, we can now write the expectation value of an observable \(O\) as
\begin{align}
\langle O \rangle
\;=\;
\frac{
\sum_{[n_i,n_b^u,n_b^d; n_O]}
U_I^{n_I} U_B^{n_B}
\det(\tilde{G}_u)\, \det(\tilde{G}_d)}
{\sum_{[n_i,n_b^u,n_b^d]}
U_I^{n_I} U_B^{n_B}
\det(G_u)\, \det(G_d)}.
\label{eq:obs-fbmc}
\end{align}
Here the numerator includes configurations with additional monomer insertions associated with the observable \(O\), collectively denoted by \(n_O\), while the denominator corresponds to the partition function. In the next section, we describe the Monte Carlo algorithm developed to evaluate \(\chi_{ud}\) and \(\chi_{uu}\) using this representation.

\section{Monte Carlo Algorithm}
\label{sec-4}

Our Monte Carlo algorithm employs three distinct update types that modify the instanton--dimer configurations $[n_i,n_b^u,n_b^d]$ contributing to the partition function in \cref{eq:pf-fb}. Each update satisfies detailed balance, but not all are ergodic individually; ergodicity is obtained by combining the three updates. The updates are: (i) the dimer update, (ii) the instanton update, and (iii) the worm update. The dimer and instanton updates add or remove dimers and instantons, respectively, and are not ergodic on their own. The worm update is ergodic but is qualitatively different: it samples extended configurations of the form $[n_i,n_b^u,n_b^d;n_O]$ that contribute to the two susceptibilities. We also introduce a tunable reweighting factor $\Omega$ for configurations in the worm sector to optimize the efficiency of the algorithm. Increasing $\Omega$ enhances the relative probability of transitions from the partition-function sector to the worm sector, thereby increasing the sampling frequency of worm configurations that contribute to the susceptibilities. This generally reduces the equilibration time measured in the number f worm updates, although each step becomes computationally more expensive because the algorithm remains in the worm sector for more time. Since $\Omega$ acts as a reweighting parameter, it is divided out when computing physical observables, leaving the expectation values unchanged.

When all the three updates are used together the overall algorithm is both ergodic and efficient. As a consistency check, we compare the results of our algorithm with exact calculations on an $L=2$ lattice in the final subsection.

\subsection{Dimer Update}
\label{sec-4a}

The purpose of the dimer update is to add or remove dimers from a given instanton--dimer configuration. The update begins by choosing, with equal probability, whether to add or remove a $u$-dimer or a $d$-dimer.

Consider the case in which a $u$-dimer is proposed to be added. For a given configuration, we count $N_F^u$, the number of free bonds in the $u$ layer at which a dimer can be added. This is done by scanning all even sites that do not have a dimer or an instanton attached to them and, for each such site, examining its six neighboring sites that also do not have a dimer or an instanton attached. Each such nearest-neighbor pair contributes one free bond to $N_F^u$. We then randomly select one of these $N_F^u$ bonds and propose to add a $u$-dimer on that bond. If the proposal is accepted, we obtain a new configuration in which the total number of $u$-dimers is denoted by $\tilde{N}_D^u$.

Let $\det(G_u^i)$ denote the determinant factor associated with the $u$ fermions in \cref{eq:pf-fb} for the initial configuration, and let $\det(G_u^f)$ denote the corresponding factor for the final configuration after the additional $u$-dimer is added. The proposal to add the $u$-dimer is accepted with
probability
\begin{align}
P_{i\rightarrow f}^{\rm acc}
= \min\!\left[
1,\;
U_B\;\frac{N_F^u}{\tilde{N}_D^u}\;
\frac{\det G_u^f}{\det G_u^i}
\right].
\label{eq:DU-itof}
\end{align}
which follows the standard Metropolis rule, with acceptance probabilities greater than unity set equal to one.

The reverse process corresponds to removing a $u$-dimer from a configuration containing an additional dimer. In this case, the proposal is accepted with probability
\begin{align}
P_{f\rightarrow i}^{\rm acc}
= \min\!\left[
1,\;
\frac{1}{U_B}\frac{\tilde{N}_D^u}{N_F^u}\;
\frac{\det G_u^i}{\det G_u^f}
\right].
\label{eq:DU-ftoi}
\end{align}
It is straightforward to verify that the acceptance probabilities in \cref{eq:DU-itof} and \cref{eq:DU-ftoi} satisfy detailed balance. The above procedure generalizes straightforwardly to the addition and removal of $d$-dimers.

\subsection{Instanton Update}
\label{sec-4b}

The purpose of the instanton update is to add or remove instantons from a given instanton--dimer configuration. To ensure that all proposed configurations have nonzero weights, instantons are always added or removed in pairs, with one instanton located on an even site and the other on an odd site. The update begins by choosing, with equal probability, whether to add or remove an instanton pair.

Consider first the case in which an instanton pair is added. For a given configuration, we count $N_F^e$ and $N_F^o$, the numbers of even and odd lattice sites, respectively, at which an instanton can be added. One can show that $N_F^e = N_F^o$. We then randomly select one even site from the $N_F^e$ possibilities and one odd site from the $N_F^o$ possibilities, and propose to add an instanton at each of these sites. If the proposal is accepted, we obtain a new configuration in which the total numbers of instantons on even and odd sites are denoted by $\tilde{N}_I^e$ and $\tilde{N}_I^o$, respectively. These quantities also satisfy $\tilde{N}_I^e = \tilde{N}_I^o$.

Let $\det(G_u^i)\det(G_d^i)$ denote the determinant factor in \cref{eq:pf-fb} for the initial configuration, and let $\det(G_u^f)\det(G_d^f)$ denote the corresponding factor for the final configuration after the instanton pair is added. The proposal to add the instanton pair is accepted with probability
\begin{align}
P_{i\rightarrow f}^{\rm acc}
= \min\!\left[
1,\;U_I^2\;
\frac{N_F^e N_F^o}{\tilde{N}_I^e \tilde{N}_I^o}\;
\frac{\det(G_u^f)\det(G_d^f)}{\det(G_u^i)\det(G_d^i)}
\right].
\label{eq:IU-itof}
\end{align}

The reverse process corresponds to removing an instanton pair from the configuration containing the additional instantons. In this case, the proposal to remove the instanton pair is accepted with probability
\begin{align}
P_{f\rightarrow i}^{\rm acc}
= \min\!\left[
1,\;\frac{1}{U_I^2}
\frac{\tilde{N}_I^e \tilde{N}_I^o}{N_F^e N_F^o}\;
\frac{\det(G_u^i)\det(G_d^i)}{\det(G_u^f)\det(G_d^f)}
\right].
\label{eq:IU-ftoi}
\end{align}
It is straightforward to verify that the acceptance probabilities in \cref{eq:IU-itof} and \cref{eq:IU-ftoi} satisfy detailed balance.

\subsection{Worm Update}
\label{sec-4c}

The worm update creates and annihilates pairs of monomers; each monomer may be either $u$-type or $d$-type. It consists of three steps: Begin, Move, and End. In the Begin step two monomers are inserted, one at the tail and one at the head of the worm. During the Move step the head monomer is displaced while the tail remains fixed. The End step annihilates the head and tail monomers.

During the update we sample extended configurations of the form $[n_i,n_b^u,n_b^d;n_O]$, which contribute directly to the observables defined in \cref{eq:obs-fbmc}. The algorithm starts in the partition-function sector, transitions to the worm sector containing the two extra monomers at the Begin step, and remains in that sector during the Move steps. The End step returns the algorithm to the partition-function sector. Thus the Begin and End steps are mutually reverse operations and together satisfy detailed balance. The Move step is composed of reversible local moves that satisfy their own detailed-balance conditions.

The weight of each worm-sector configuration is determined by the observable it samples up to the reweighting factor $\Omega$. Hence, averaging the number of worm steps that contribute to a given observable yields its expectation value up to the same reweighting factor. For example, worm configurations with a $u$-type monomer at the head and a $d$-type monomer at the tail contribute to $\chi_{ud}$, while configurations with two $u$-type monomers at those locations contribute to $\chi_{uu}$. We describe each step in detail below.

\subsubsection{Begin--End Step}

The Begin step initiates the worm update by selecting a lattice site $i$ at random. With equal probability, this site lies on either the $u$ layer or the $d$ layer. During the Begin step, the configuration transitions from the partition-function sector to the worm sector through the creation of two monomers, identified as the Tail and the Head. The Tail is fixed at the original site $i$, while the Head is placed on a neighboring site $j$.

\begin{figure}
\centering
\includegraphics[width=0.35\textwidth]{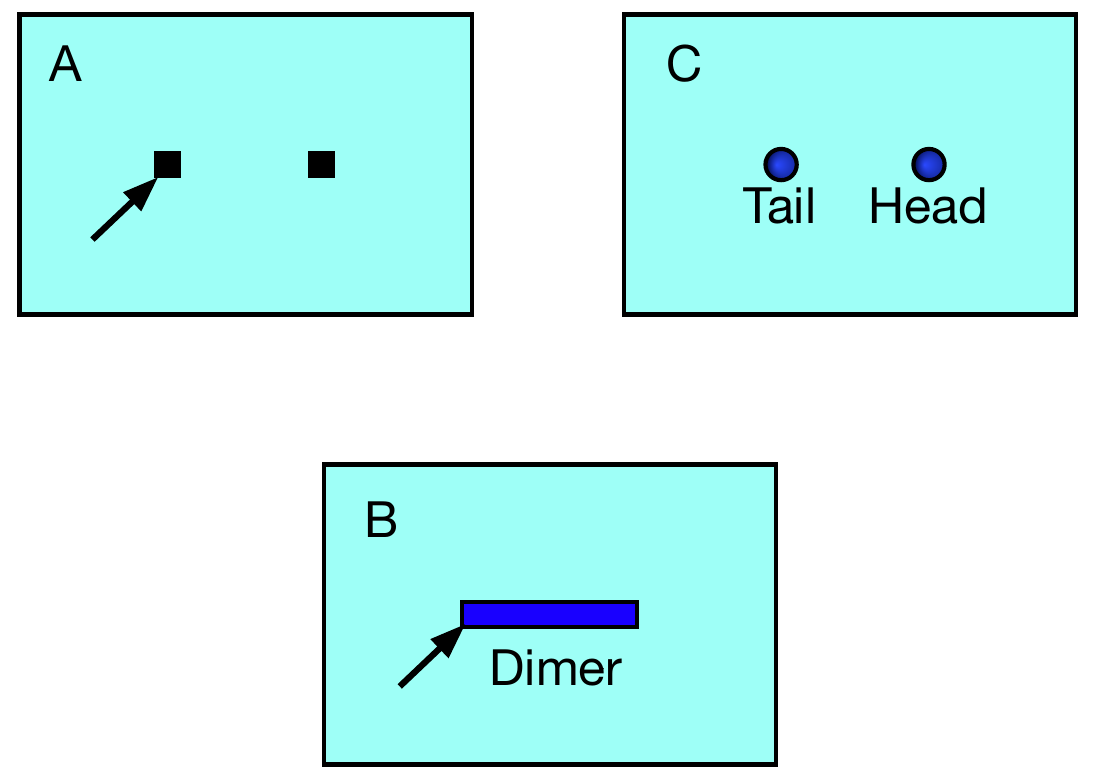}
\caption{Begin--End step of the worm algorithm restricted to a single layer. During the Begin step, the initially chosen site may be either a free site (configuration $A$) or a site occupied by a dimer (configuration $B$). The solid arrow indicates the site selected at the start of the Begin step. Configurations $A$ and $B$ belong to the partition-function sector. From either configuration, a proposal is made to transition to configuration $C$ in the worm sector, with the Tail located at the initially chosen site and the Head placed on a neighboring site. For configuration $A$, the Head is chosen at random from the $\sD$ nearest-neighbor sites on the same layer, while for configuration $B$ it is placed on the fixed neighboring site shown in the figure. The worm update may also terminate immediately, transitioning directly from configuration $A$ to $B$ or from $B$ to $A$. If the Begin step produces configuration $C$, the End step removes the worm by transforming configuration $C$ back into either configuration $A$ or $B$. The same configurations appear in the End step as in the Begin step, but with the arrow indicating the reverse direction of the update. In this figure, both the Tail and the Head are created on the same fermion layer, either the $u$ layer or the $d$ layer.}
\label{fig:begin-end-dimer}
\end{figure}

We first consider the case in which the initial site $i$ is either free (i.e., contains neither an instanton nor a dimer) or occupied by a dimer. This situation is illustrated in \cref{fig:begin-end-dimer}, where the arrow indicates the initially selected site $i$. Configuration~A corresponds to the case in which the site is free, while configuration~B corresponds to the case in which the site contains a dimer.

Starting from configuration~A, one of the $\sD = 6$ neighboring sites of site $i$ on the same layer is chosen at random and denoted by $j$. If site $j$ contains either a dimer or an instanton, the worm update terminates. Otherwise, with equal probability, one of two proposals is made: either a dimer is created between sites $i$ and $j$, leading to configuration~B, or a worm configuration is created with a Tail monomer at site $i$ and a Head monomer at site $j$, leading to configuration~C shown in \cref{fig:begin-end-dimer}.

The proposal to transition from configuration~A to configuration~B is accepted with probability
\begin{align}
P_{A \rightarrow B}
= \min\!\left[
1,\;
U_B\,\sD \,\frac{\det G^B}{\det G^A}
\right],
\end{align}
while the proposal to transition from configuration~A to configuration~C is accepted with probability
\begin{align}
P_{A \rightarrow C}
= \min\!\left[
1,\;
\frac{\Omega\,\sD}{\sD+1}\,\frac{\det G^C}{\det G^A}
\right].
\end{align}
In both expressions, $\det G^A$ denotes the fermion determinant factor for configuration~A, while $\det G^B$ denotes the corresponding factor for the proposed configuration~B, which includes the additional dimer between $i$ and $j$. Similarly, $\det G^C$ is the determinant factor with an additional monomers at the Head and Tail sites. The factor $\Omega$ is the reweighting factor of the worm sector.

If instead the Begin step starts from configuration~B, we propose, with equal probability, a transition either to configuration~A or to configuration~C. The proposal to transition from configuration~B to configuration~A is accepted with probability
\begin{align}
P_{B \rightarrow A}
= \min\!\left[
1,\;
\frac{1}{U_B\,\sD}\,\frac{\det G^A}{\det G^B}
\right],
\end{align}
while the proposal to transition from configuration~B to configuration~C is accepted with probability
\begin{align}
P_{B \rightarrow C}
= \min\!\left[
1,\;
\frac{\Omega}{U_B\,(\sD+1)}
\right].
\end{align}

The End step corresponds to transitions from configuration~C back to either configuration~A or configuration~B. At the start of this step, it is not known a priori whether the update will result in a Move step or an End step. One of the $\sD+1$ neighboring sites of the Head is chosen at random; these include the $\sD$ nearest-neighbor sites on the same layer, as well as the site on the opposite layer with the same space--time coordinates. If the selected neighboring site coincides with the Tail site on the same layer, the step is identified as an End step, and a proposal is made to transition to either configuration~A or configuration~B.

The proposal to transition from configuration~C to configuration~A is accepted with probability
\begin{align}
P_{C \rightarrow A}
= \min\!\left[
1,\;
\frac{\sD+1}{\Omega\,\sD}\,\frac{\det G^A}{\det G^C}
\right],
\end{align}
while the proposal to transition from configuration~C to configuration~B is accepted with probability
\begin{align}
P_{C \rightarrow B}
= \min\!\left[
1,\;
\frac{U_B\,(\sD+1)}{\Omega}
\right].
\end{align}
According to the standard Metropolis prescription, if a proposed update is rejected, the original configuration is retained as the new configuration.

\begin{figure}
\centering
\includegraphics[width=0.15\textwidth]{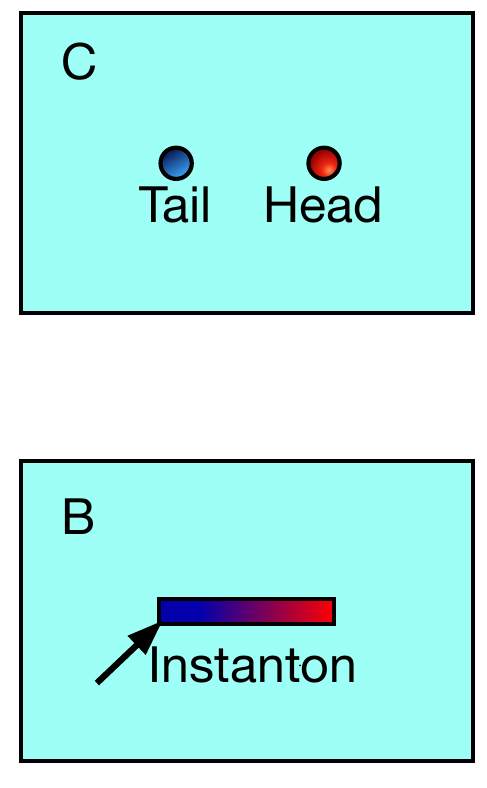}
\caption{Begin--End step of the worm algorithm involving opposite layers. During this Begin step, the initially selected site contains an instanton (configuration~B). The solid arrow indicates the site chosen at the start of the Begin step. Configuration~B belongs to the partition-function sector. A proposal is then made to transition to configuration~C in the worm sector, with the Tail located at the initially selected site and the Head placed on the site with the same space--time coordinates on the opposite layer. If the proposal is accepted, the instanton is broken into two monomers. The End step is the reverse process, in which the two monomers on the two layers are fused to form an instanton.}
\label{fig:begin-end-instanton}
\end{figure}

The Begin step can also occur when the initially selected site $i$ is connected to an instanton, as illustrated in \cref{fig:begin-end-instanton}. In this case, the transition to the worm sector proceeds by breaking the instanton into two monomers, with the Head placed on the site with the same space--time coordinates as $i$ but on the opposite layer. The resulting worm-sector configuration is denoted as configuration~C in \cref{fig:begin-end-instanton}.

For this process, the proposal to transition from configuration~B to configuration~C is accepted with probability
\begin{align}
P_{B \rightarrow C}
= \min\!\left[
1,\;
\frac{\Omega}{U_I\,(\sD+1)}
\right],
\end{align}
while the reverse proposal to transition from configuration~C back to configuration~B is accepted with probability
\begin{align}
P_{C \rightarrow B}
= \min\!\left[
1,\;
\frac{U_I\,(\sD+1)}{\Omega}
\right].
\end{align}

\begin{figure}
\centering
\includegraphics[width=0.35\textwidth]{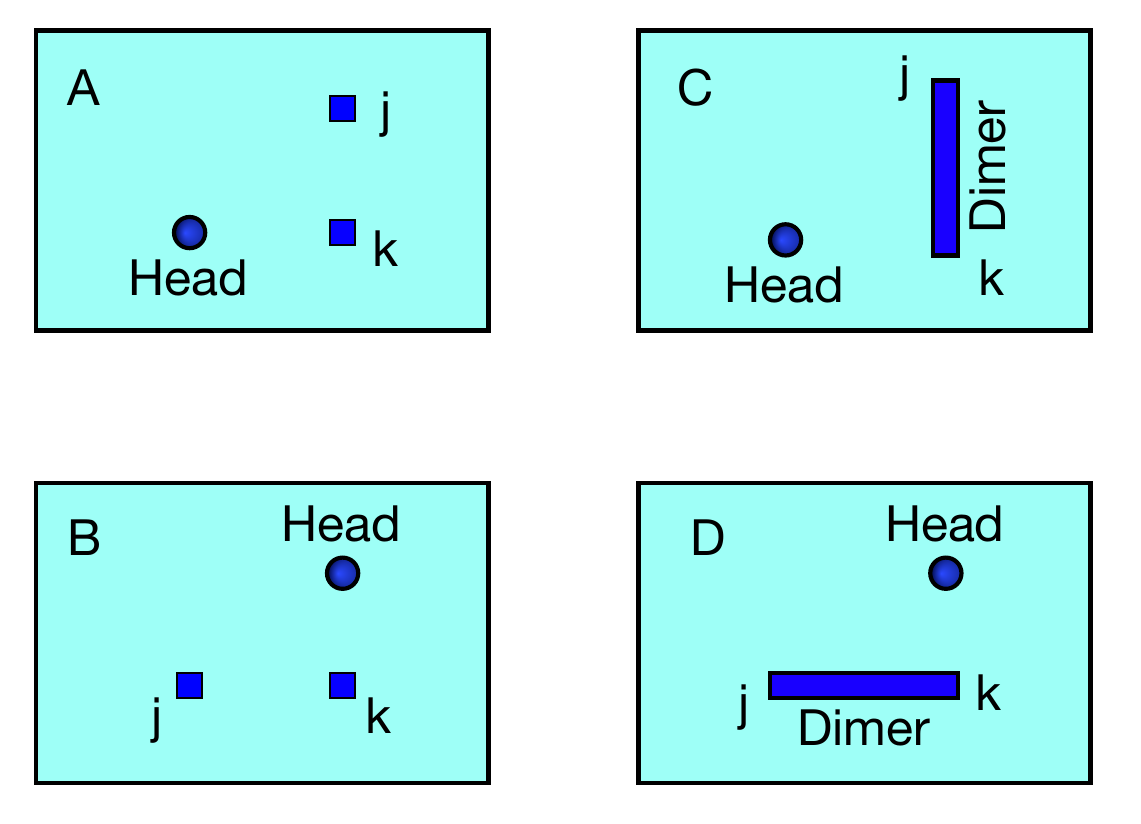}
\caption{First class of configurations among which the Move step transitions. In this class, the chosen neighboring site $k$ of the Head lies on the same layer and is either a free site or a site occupied by a dimer. If $k$ is free, the site $j$ is chosen at random from the remaining nearest-neighbor sites on the same layer, excluding the Head site.}
\label{fig:move-dimer}
\end{figure}

\begin{figure}
\centering
\includegraphics[width=0.35\textwidth]{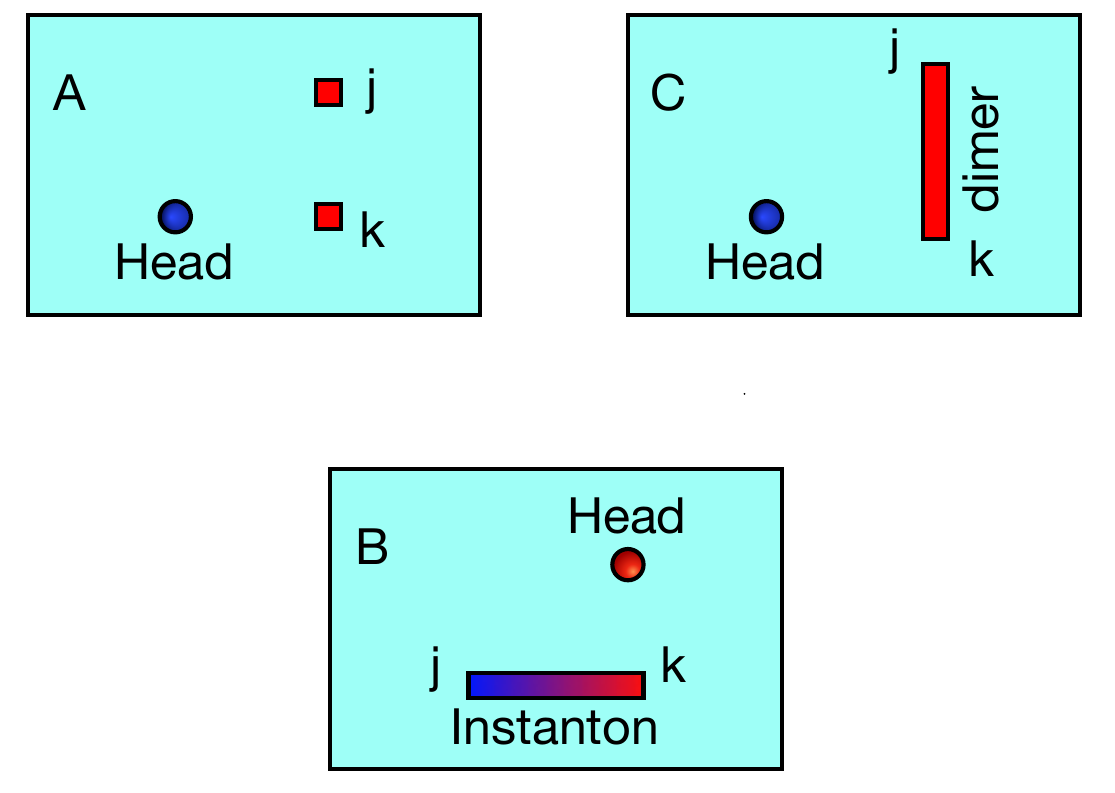}
\caption{Second class of configurations among which the Move step transitions. In this class, the chosen neighboring site $k$ of the Head lies on the opposite layer and is either a free site or a site occupied by a dimer. The site $k$ may also lie on the same layer as the Head if it is occupied by an instanton. If $k$ is free, the site $j$ is chosen at random from the nearest-neighbor sites on the same layer.}
\label{fig:move-instanton}
\end{figure}

\subsubsection{Move Step}

During the Move step, the Head monomer is displaced to a next-to-nearest-neighbor site. Specifically, one of the $(\sD+1)$ neighboring sites of the Head is chosen at random and labeled as site $k$. These neighboring sites consist of the $\sD$ nearest neighbors on the same layer, together with the site on the opposite layer that has the same space--time coordinates.

Depending on the configuration at site $k$, the Move step is divided into two classes of updates, illustrated in \cref{fig:move-dimer} and \cref{fig:move-instanton}. The first class consists of configurations in which site $k$ lies on the same layer as the Head and is either a free site (i.e., not connected to an instanton or a dimer) or a site occupied by a dimer. If $k$ is a free site, we select one of its $\sD$ neighboring sites on the same layer, including the current Head site, and label it as site $j$. If $k$ contains a dimer, the site connected to $k$ by that dimer is uniquely identified as site $j$ and lies on the same layer. This class contains the four configurations shown in \cref{fig:move-dimer}.

The second class consists of configurations in which site $k$ either lies on the opposite layer and is a free site, or lies on the same layer as the Head and contains an instanton. If $k$ is a free site on the opposite layer, we choose one of its $\sD$ neighboring sites on that layer and label it as site $j$. If $k$ is on the same layer and contains an instanton, the site connected to $k$ through the instanton, which lies on the opposite layer, is uniquely identified as site $j$. In this case, the Head and site $j$ reside on opposite layers. This class contains the three configurations shown in \cref{fig:move-instanton}.

In both classes of configurations, if site $k$ is free but site $j$ contains either a dimer or an instanton, or is the same as the Head site, the Head does not move and remains at its current location. Such configurations are not shown in the figures. Otherwise, the Head is displaced according to the Metropolis acceptance probabilities discussed below, which depend on whether the current configuration belongs to the first or the second class.

Within the first class of configurations, consider the case in which site $k$ is free. The initial configuration may then be identified with configuration~A in \cref{fig:move-dimer}. A proposal is made to transition, with equal probability, to one of the remaining three configurations shown in \cref{fig:move-dimer}. These proposals are accepted according to the Metropolis probabilities
\begin{align}
P_{A \rightarrow B}
&= \min\!\left[
1,\;
\frac{\det G^B}{\det G^A}
\right], \\
P_{A \rightarrow C}
&= \min\!\left[
1,\;
U_B\,\sD\,\frac{\det G^C}{\det G^A}
\right], \\
P_{A \rightarrow D}
&= P_{A \rightarrow C}.
\end{align}
Here, $\det G^A$ denotes the fermion determinant factor in the initial configuration, which contains a monomer at the Head site. The factor $\det G^B$ denotes the corresponding determinant when the Head monomer is moved to site $j$, while $\det G^C$ denotes the determinant factor for the configuration containing two additional monomers at sites $k$ and $j$, in addition to the monomer at the Head site. It is useful to recognize that the determimant factor $\det G^D$ for confguration~D is the same as $\det G^C$.

Within the first class of configurations, if site $k$ contains a dimer, the initial configuration may be identified with configuration~C in \cref{fig:move-dimer}. In this case, with equal probability, a proposal is made to transition to configuration~A, configuration~B, or configuration~D. The corresponding acceptance probabilities are
\begin{align}
P_{C \rightarrow A}
&= \min\!\left[
1,\;
\frac{1}{U_B\,\sD}\,\frac{\det G^A}{\det G^C}
\right], \\
P_{C \rightarrow B}
&= \min\!\left[
1,\;
\frac{1}{U_B\,\sD}\,\frac{\det G^B}{\det G^C}
\right], \\
P_{C \rightarrow D}
&= 1.
\end{align}

If site $k$ lies on the opposite layer and is a free site, the initial configuration may be identified with configuration~A in the second class of configurations shown in \cref{fig:move-instanton}. In this case, a proposal is made, with equal probability, to transition to configuration~B or configuration~C. These proposals are accepted with probabilities
\begin{align}
P_{A \rightarrow B}
&= \min\!\left[
1,\;
U_I\,\sD\,\frac{\det G^B}{\det G^A}
\right], \\
P_{A \rightarrow C}
&= \min\!\left[
1,\;
U_B\,\sD\,\frac{\det G^C}{\det G^A}
\right].
\end{align}
Here, as before, $\det G^A$ denotes the fermion determinant factor for configuration~A, which contains a monomer at the Head site. The factor $\det G^B$ denotes the corresponding determinant for the configuration~B in which two additional monomers are added at sites $k$ and $j$ on the opposite layer. The determinant factor $\det G^C$ for the configuration~C is the same as $\det G^B$.

If site $k$ lies on the opposite layer and is connected to a dimer, the initial configuration may be identified with configuration~C in the second class of configurations shown in \cref{fig:move-instanton}. In this case, a proposal is made, with equal probability, to transition to configuration~A or configuration~B. These proposals are accepted with probabilities
\begin{align}
P_{C \rightarrow A}
&= \min\!\left[
1,\;
\frac{1}{U_B\,\sD}\,\frac{\det G^A}{\det G^C}
\right], \\
P_{C \rightarrow B}
&= \min\!\left[
1,\;
\frac{U_I}{U_B}
\right].
\end{align}

If instead site $k$ lies on the same layer as the Head and is connected to an instanton, the initial configuration may be identified with configuration~B in the second class of configurations shown in \cref{fig:move-instanton}. In this case, a proposal is made, with equal probability, to transition to configuration~A or configuration~C. These proposals are accepted with probabilities
\begin{align}
P_{B \rightarrow A}
&= \min\!\left[
1,\;
\frac{1}{U_I\,\sD}\,\frac{\det G^A}{\det G^B}
\right], \\
P_{B \rightarrow C}
&= \min\!\left[
1,\;
\frac{U_B}{U_I}
\right].
\end{align}

The worm update is ergodic by itself and also provides an efficient way to compute the susceptibilities $\chi_{ud}$ and $\chi_{uu}$. For example, the susceptibility $\chi_{ud}$ is obtained by averaging the number of worm configurations encountered during a worm update in which the Head and Tail reside on opposite layers, and then dividing by the reweighting factor $\Omega$. Similarly, $\chi_{uu}$ is obtained by averaging the number of worm configurations encountered during a worm update for which the Head and Tail are on the same layer, again divided by the reweighting factor $\Omega$.

\subsection{Idea of Fluctuation Matrices}
\label{sec-4d}

One of the most time-consuming steps in the fermion-bag Monte Carlo algorithm, as in other fermion Monte Carlo methods \cite{ALF:2020tyi}, is the calculation of ratios of determinants that appear in many of the Metropolis accept/reject probabilities discussed above. It is well known in the fermion-algorithm literature that computing individual determinants and then taking their ratio is both numerically unstable and computationally inefficient. A more robust approach is to compute the ratio of determinants directly. This is particularly advantageous when the two determinants differ only by small, local changes in the fermion configuration.

In the fermion-bag approach, such ratios of determinants naturally arise from local fluctuations of fermion bags about a chosen background configuration. These ratios can be reformulated as the determinant of a much smaller matrix, referred to as a fluctuation matrix. This idea was explained in detail in \cite{Ayyar:2016xwv} and is closely related to standard results from linear algebra. Consider two large square matrices $G$ and $G_B$, where $G$ is obtained from a base matrix $G_B$ by deleting $k_d$ rows and adding $k_a$ rows. One can then show that
\begin{align}
\frac{\det G}{\det G_B} = \det F,
\end{align}
where $F$ is a square matrix of dimension $k = k_d + k_a$. The matrix $F$ is constructed from the matrix elements corresponding to the added and removed rows, together with the appropriate elements of $G_B^{-1}$. In this formulation, the effect of the local fluctuation is entirely encoded in the small matrix $F$, whose elements depend on the large-scale structure of the fermion bag through the inverse of the base matrix $G_B^{-1}$.

This perspective allows all ratios of the form
\begin{align}
\frac{\det G_1}{\det G_2}
=
\frac{\det F_1}{\det F_2}
\end{align}
to be computed efficiently, where $F_1$ and $F_2$ are fluctuation matrices constructed with respect to a common base matrix $G_B$. Since the dimensions of the fluctuation matrices depend only on the size of the local updates, their determinants can be evaluated rapidly, leading to a significant reduction in computational cost.

In our algorithm, we ensure that the size of the fluctuation matrices remains bounded during the Monte Carlo evolution. When the size grows beyond a prescribed threshold, we reset the base matrix $G_B$ by recomputing it from the current configuration. The most time-consuming step in this procedure is precisely this resetting of the base matrix; however, it occurs infrequently and does not affect the overall efficiency of the algorithm.

We have validated our algorithm by comparing results obtained from Monte Carlo simulations with exact results on an $L=2$ lattice. In the appendix we discuss the comparison of our Monte Carlo data with exact results.

\section{Monte Carlo Results}
\label{sec-5}

In this section, we present and analyze Monte Carlo data generated with the algorithm introduced in \cref{sec-4} to determine the critical behavior and phase diagram of the model. Simulations were performed on cubic lattices with linear sizes in the range $L=8-56$. For each parameter set, simulations were preceded by typically $10^5$ Monte Carlo steps for thermalization. During thermalization, we employed a reweighting factor $\Omega=5$ for lattices with $L<40$ and $\Omega=20$ for larger lattices ($L=40-56$). This choice was motivated by explicit tests with $\Omega=1,5,$ and $20$, which showed that $\Omega=20$ improved equilibration for larger systems by effectively reducing the Monte Carlo step size.

After thermalization, measurements were accumulated over multiple statistically independent replicas. In the vicinity of the Gross-Neveu transition, lattice sizes $L=28-56$ were studied using between $10$ and $50$ replicas, depending on $L$. For lattices $L=28–36$, a total statistics of approximately $4 \times 10^6$ Monte Carlo steps was accumulated from 10 replicas for each lattice size. For larger lattices ($L=40–56)$, total statistics ranged from approximately $2.5 \times 10^6$ to $8 \times 10^6$ Monte Carlo steps, accumulated using between 20 and 50 replicas, depending on the lattice size and coupling. 
Integrated autocorrelation times were monitored for all measured observables. We quote the autocorrelation time of the instanton density, since it exhibited the largest integrated autocorrelation time among all observables. The maximum autocorrelation time for this observable was found to be at $L=52$ for couplings $U_I=0.6–0.615$. The corresponding autocorrelation time was found to be $\tau \approx 4000$. This value was obtained using the total statistics of approximately $8\times10^6$ Monte Carlo steps accumulated over multiple independent replicas, which corresponds to approximately $N_{\mathrm{ind}} \simeq N_{\mathrm{MC}} /2\tau \approx 10^3$ statistically independent measurements.

For the three-dimensional XY transition, simulations were performed for lattice sizes $L=12–32$ using $10–30$ replicas, with total statistics ranging from approximately $2 \times 10^6$ to $4 \times 10^6$ Monte Carlo steps. For lattice sizes close to the critical region, autocorrelation times grow substantially. In particular, for $L = 16$ near the critical point, we find $\tau \approx 400$ when measurements are recorded after every 10 Monte Carlo steps, whereas for larger lattices such as $L = 32$, where measurements are recorded every 2 Monte Carlo steps, autocorrelation times increase to values as large as $\tau \approx 1500$. For the largest lattice sizes and couplings closest to the transition, this results in increased statistical fluctuations, reflecting the combined effects of critical slowing down and the finite statistics available.

The statistical uncertainties quoted throughout this section were estimated using the jackknife method and were cross-checked using both bootstrap resampling and autocorrelation-corrected error analyses to ensure consistency. The discussion is organized into several subsections, each focusing on a specific aspect of the numerical results.

\begin{figure*}
\centering
\includegraphics[width=0.3\textwidth]{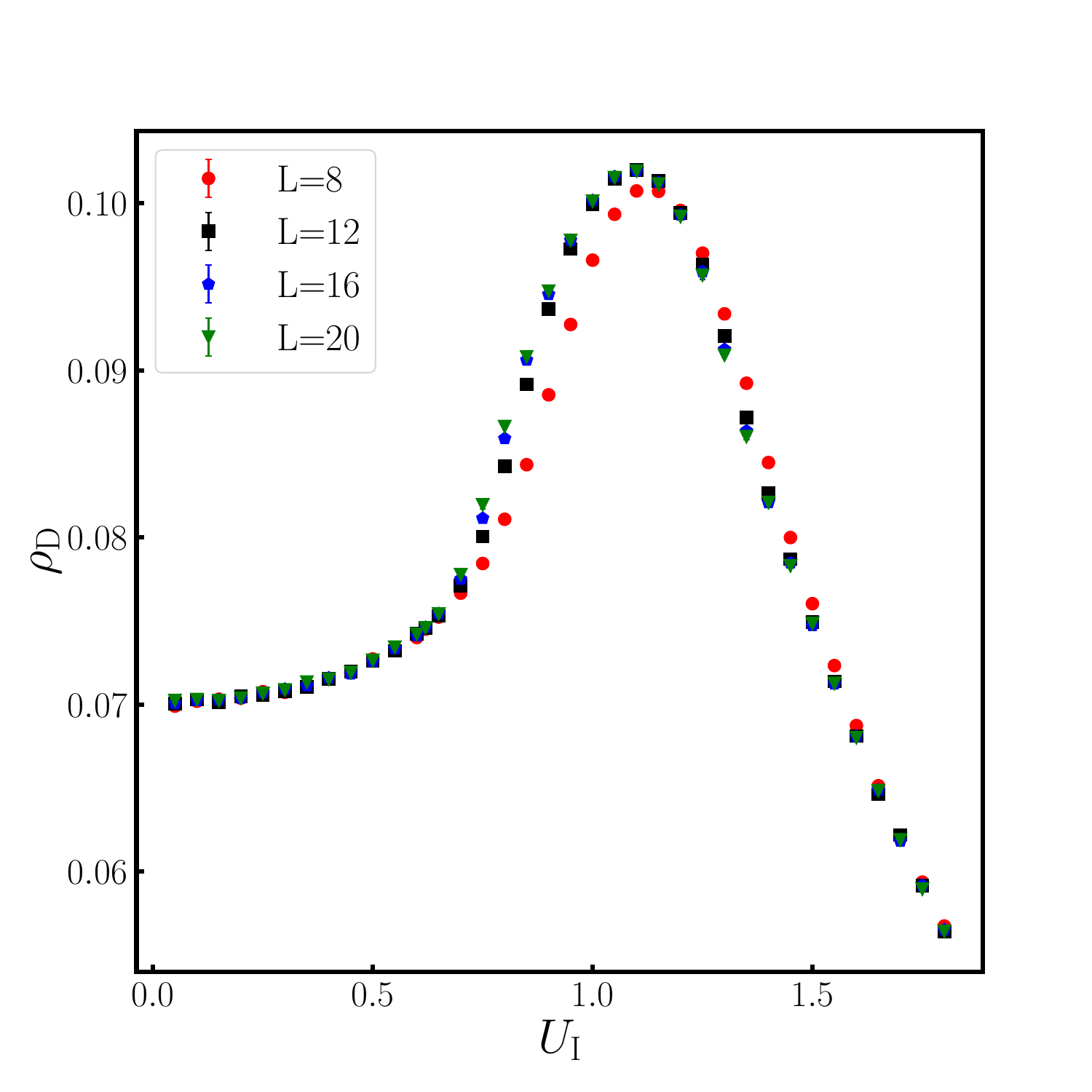}
\includegraphics[width=0.3\textwidth]{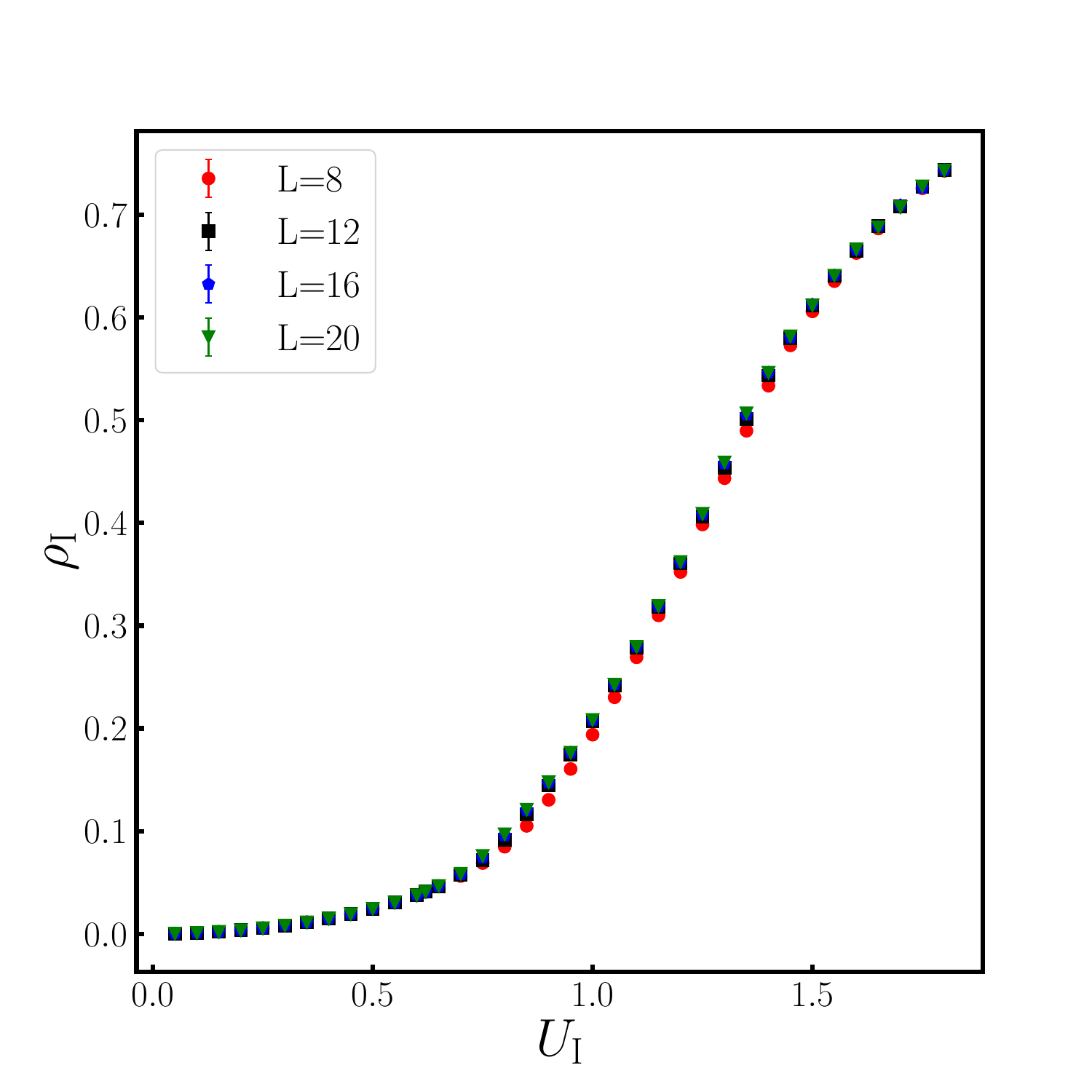}
\includegraphics[width=0.3\textwidth]{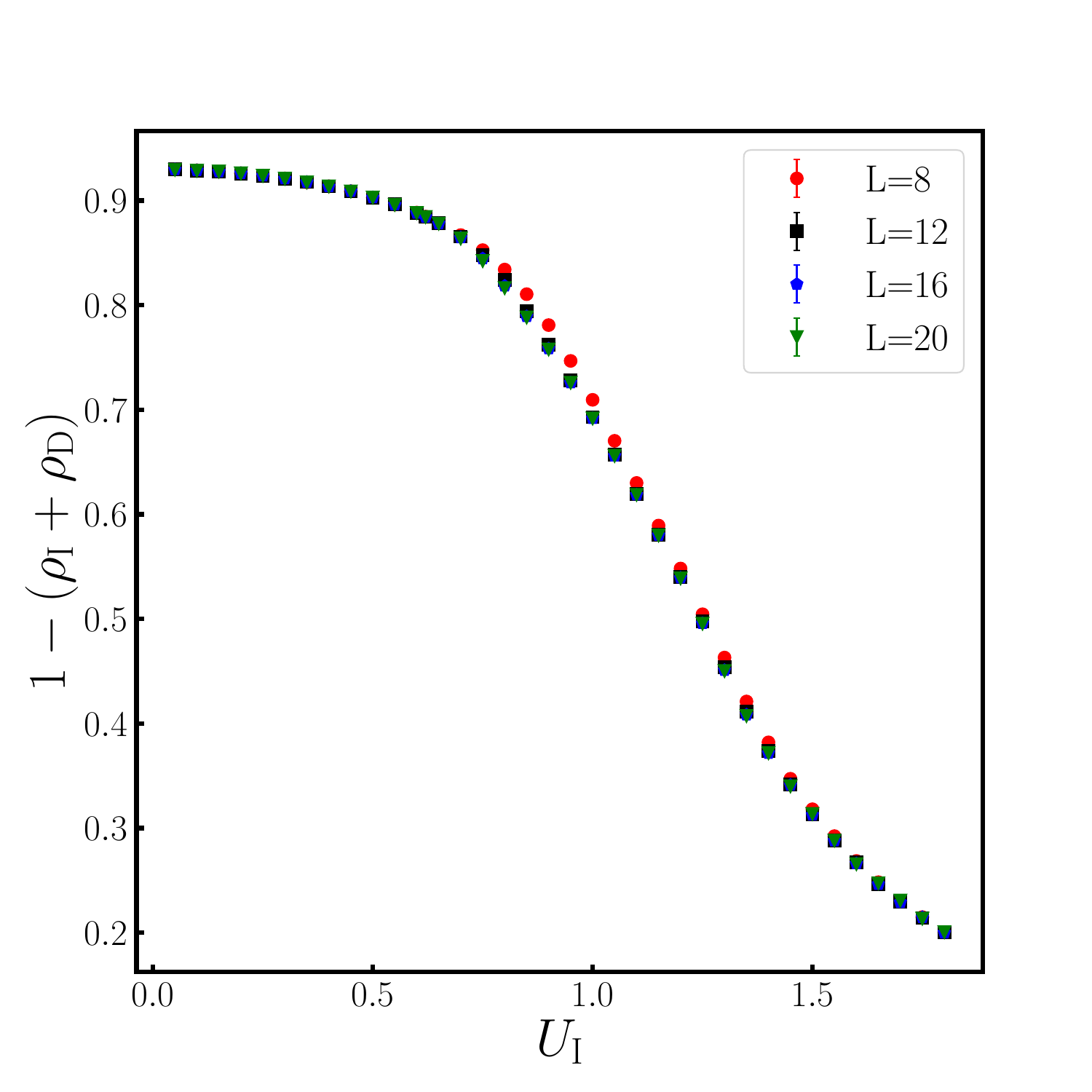}
\caption{The variation of the average dimer density ($\rho_D)$ (left panel), the average instanton density ($\rho_I$) (central panel) and the density of free sites ($1-\rho_I-\rho_D$) (right panel) as a function of $U_I$ at a fixed values of $U_B=0.1$.}
\label{fig:rho}
\end{figure*}

\subsection{Dimers, Instantons, and Free Sites}
\label{sec-5a}

The average numbers of dimers, instantons, and free sites in the Monte Carlo configurations are conveniently characterized by their corresponding densities. The dimer density $\rho_D$ and the instanton density $\rho_I$, defined in \cref{eq:obs-rhoD,eq:obs-rhoI}, provide useful measures of the strengths of the interactions $U_B$ and $U_I$. By construction, these densities are normalized to lie between $0$ and $1$. We also define the free-site density,
\begin{equation}
\rho_F = 1 - \rho_D - \rho_I,
\label{eq:obs-rhoF}
\end{equation}
which represents the fraction of lattice sites that are not connected to either a dimer or an instanton. In \cref{fig:rho}, we plot $\rho_D$, $\rho_I$, and $\rho_F$ as functions of $U_I$ at fixed $U_B = 0.1$.

As $U_I$ increases, the instanton density $\rho_I$ rises monotonically, as expected. In contrast, the dimer density $\rho_D$ exhibits a nontrivial structure: it increases rapidly at small $U_I$, reaches a maximum around $U_I \approx 1.1$, and then decreases at larger values of $U_I$. The behavior of $\rho_F$ is directly relevant for the computational cost at large $U_I$. In our simulations we employ the weak-coupling viewpoint of the fermion-bag formulation, in which the sizes of the fermion matrices $D_u$ and $D_d$, whose determinants must be evaluated, scale with the combined density $(\rho_I + \rho_D)$ (see \cref{eq:pf-1}). Consequently, when $\rho_F \lesssim 0.5$, simulations on large lattice volumes become increasingly demanding. 

For example, at $U_I \approx 1.8$ we find $\rho_F \approx 0.2$, indicating that nearly $80\%$ of lattice sites are occupied by either dimers or instantons. In this regime it would be more efficient to adopt the strong-coupling viewpoint of the fermion-bag formulation~\cite{Chandrasekharan:2013rpa}. Since the physics at such large couplings is not the focus of the present work, we restrict ourselves to the weak-coupling viewpoint, which is better suited for the parameter regime of interest.

\begin{figure*}
\centering
\hbox{
\includegraphics[width=0.3\textwidth]{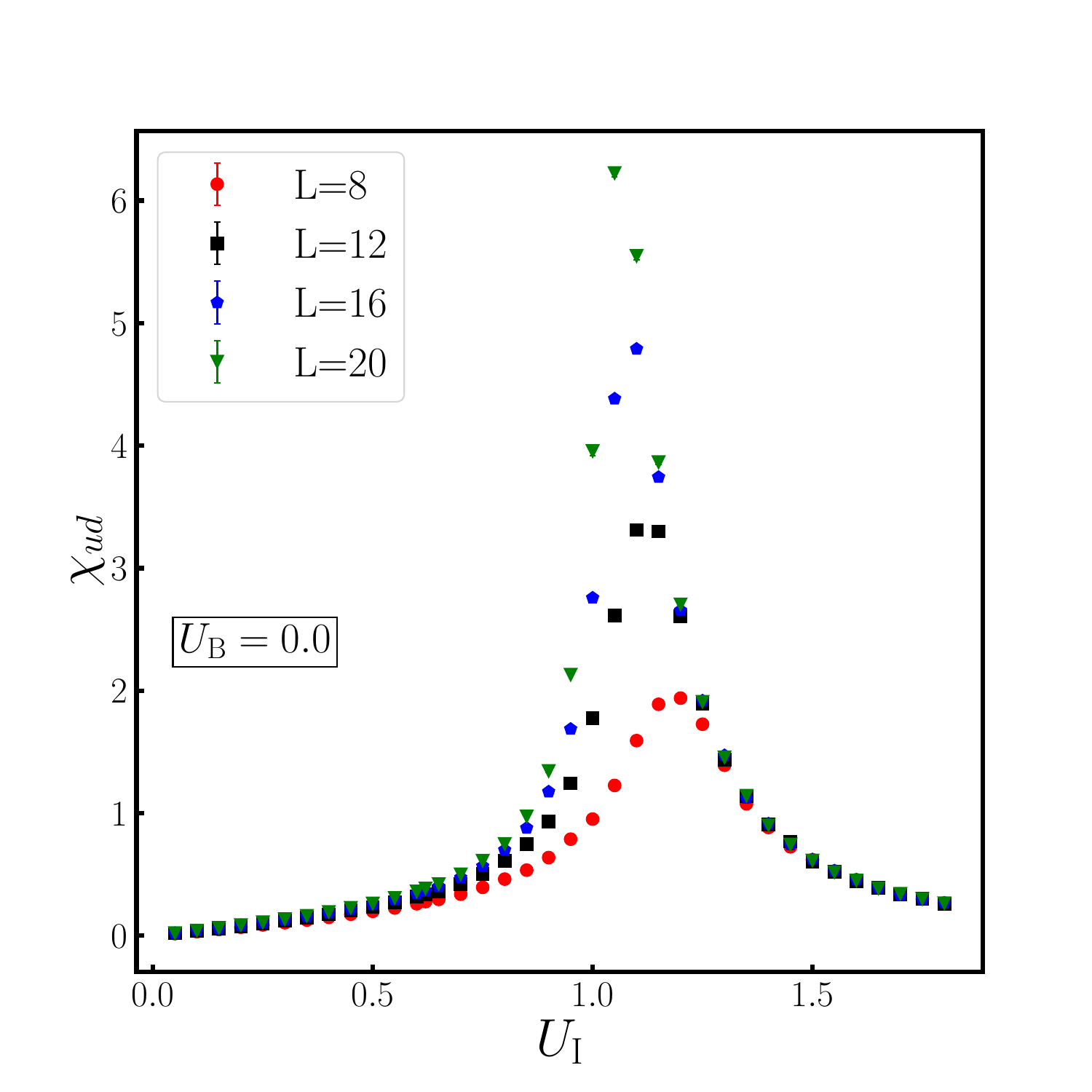}
\includegraphics[width=0.3\textwidth]{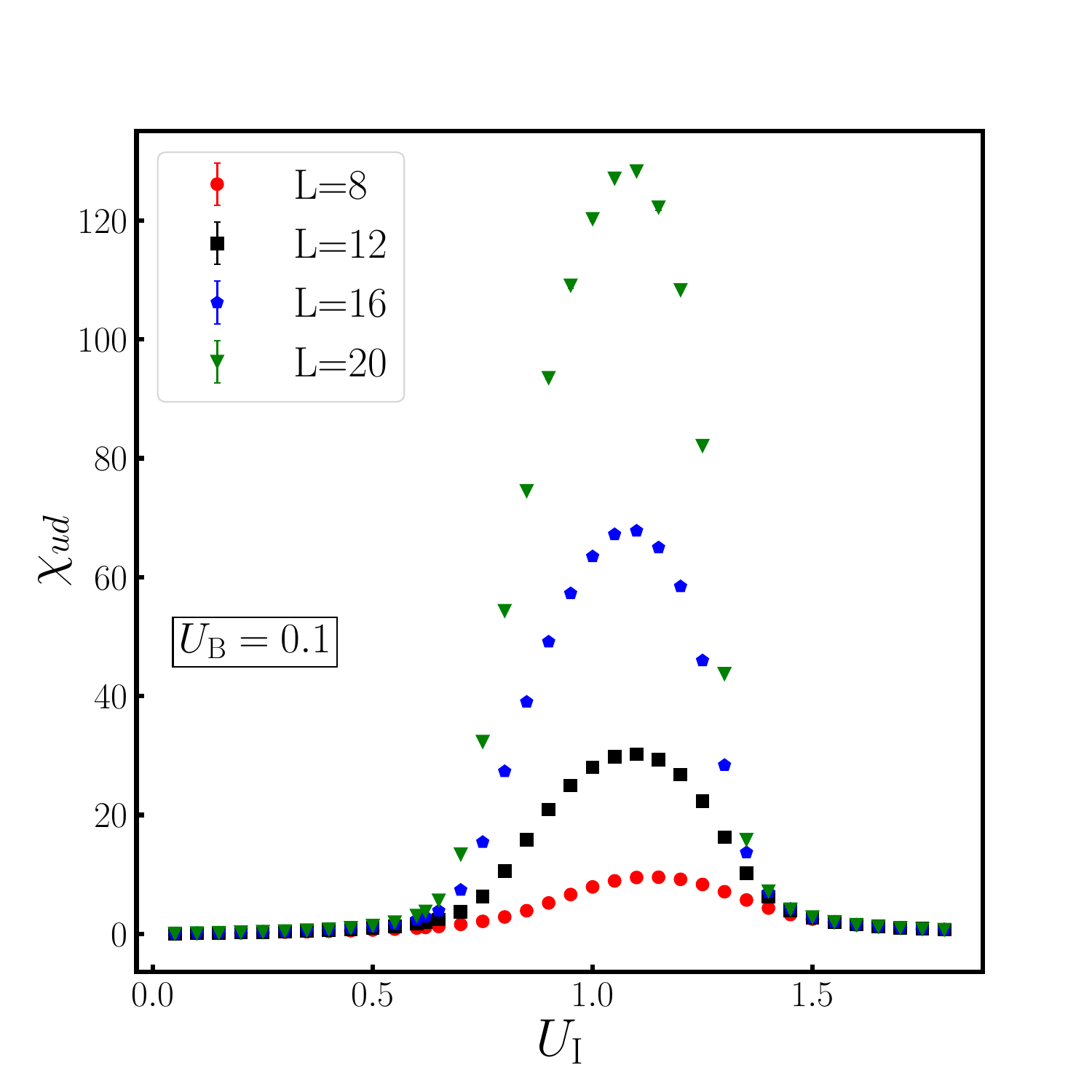}
\includegraphics[width=0.3\textwidth]{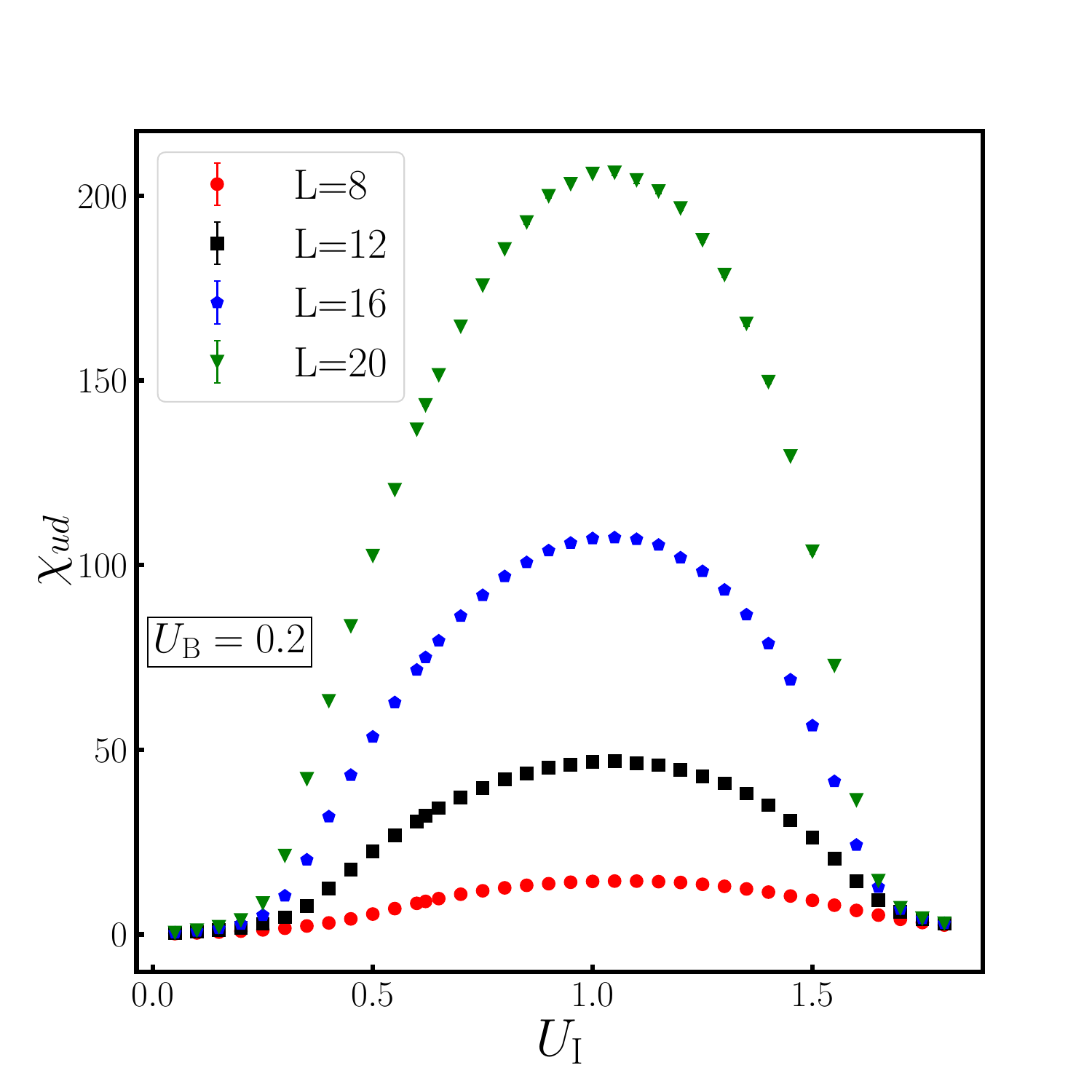}
}
\hbox{
\includegraphics[width=0.3\textwidth]{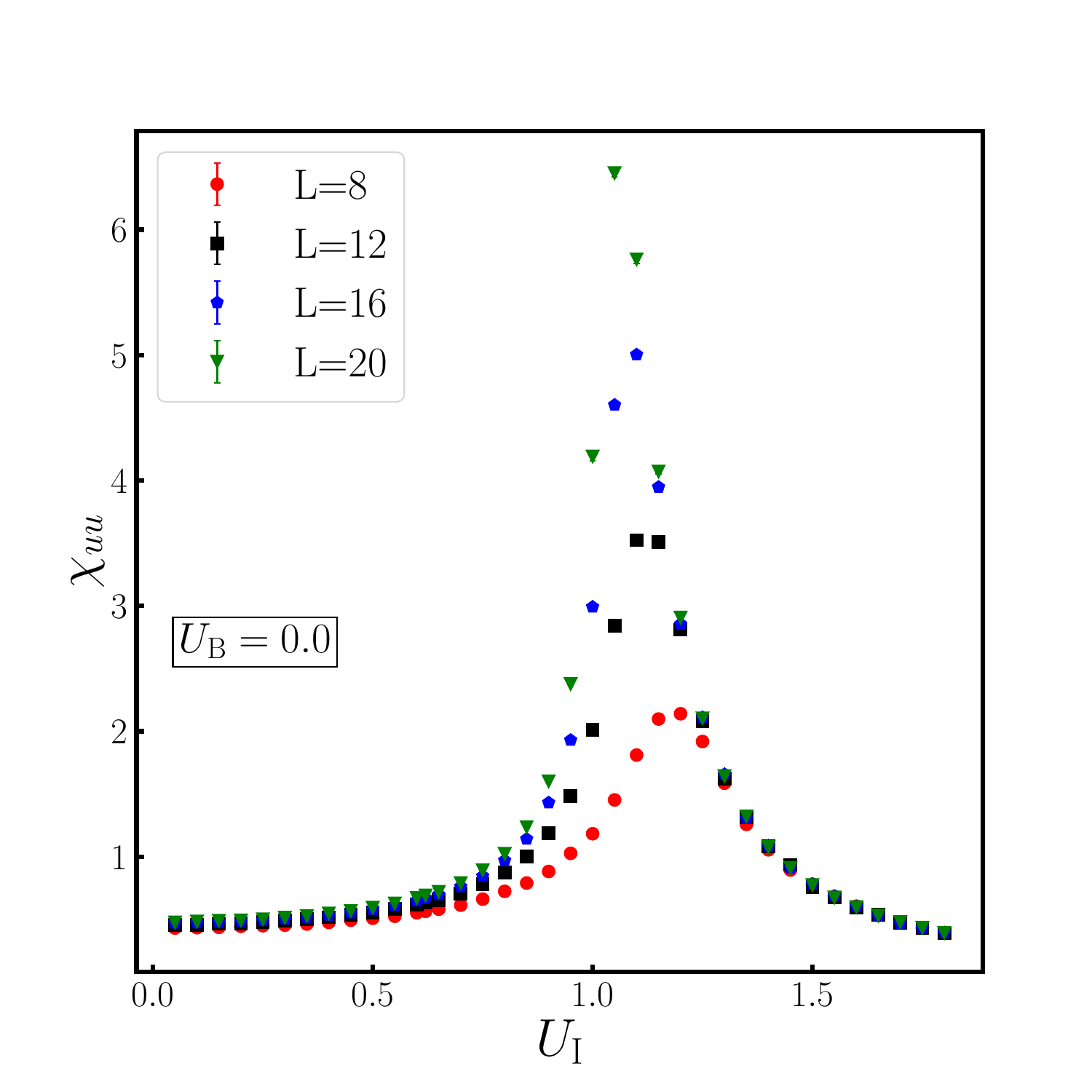}
\includegraphics[width=0.3\textwidth]{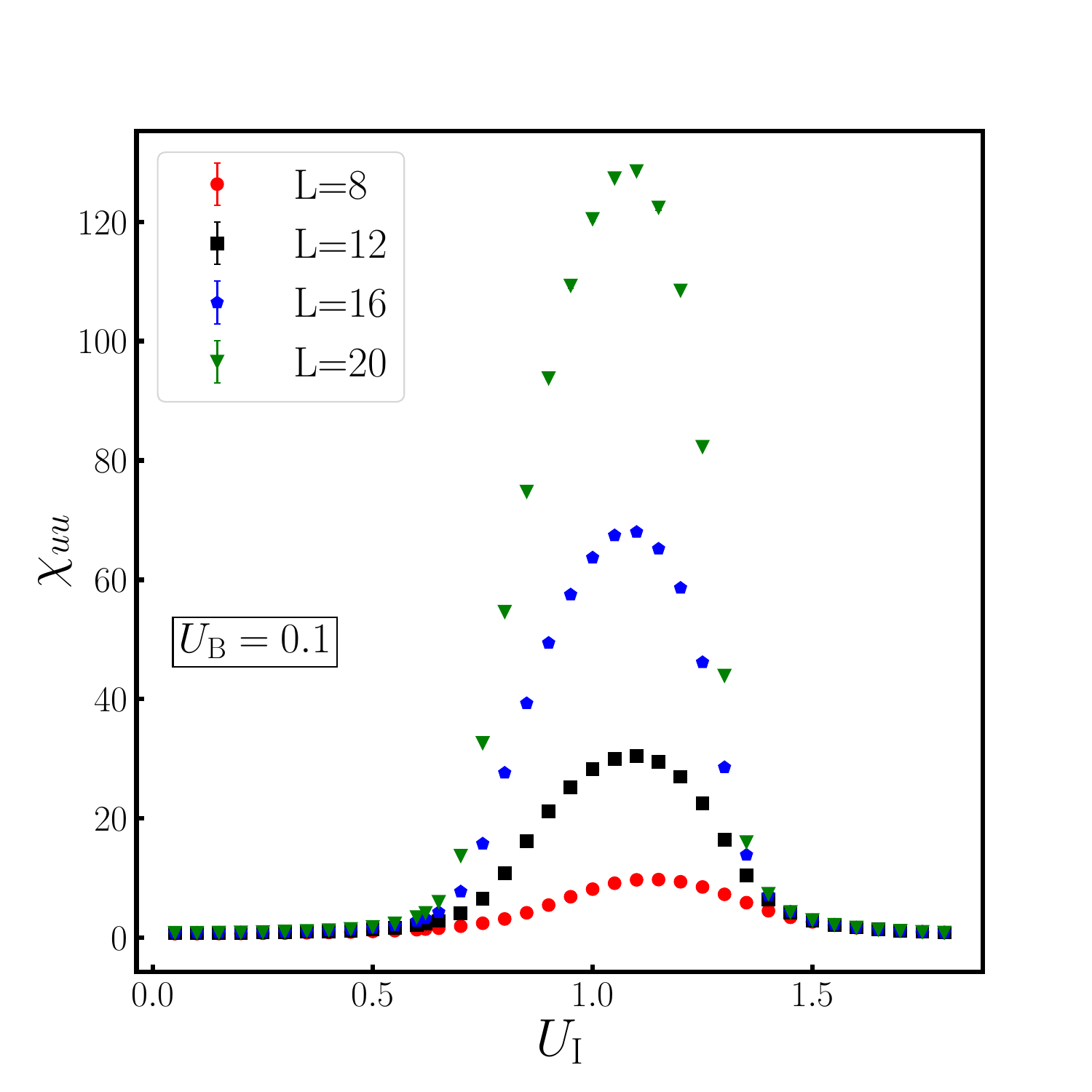}
\includegraphics[width=0.3\textwidth]{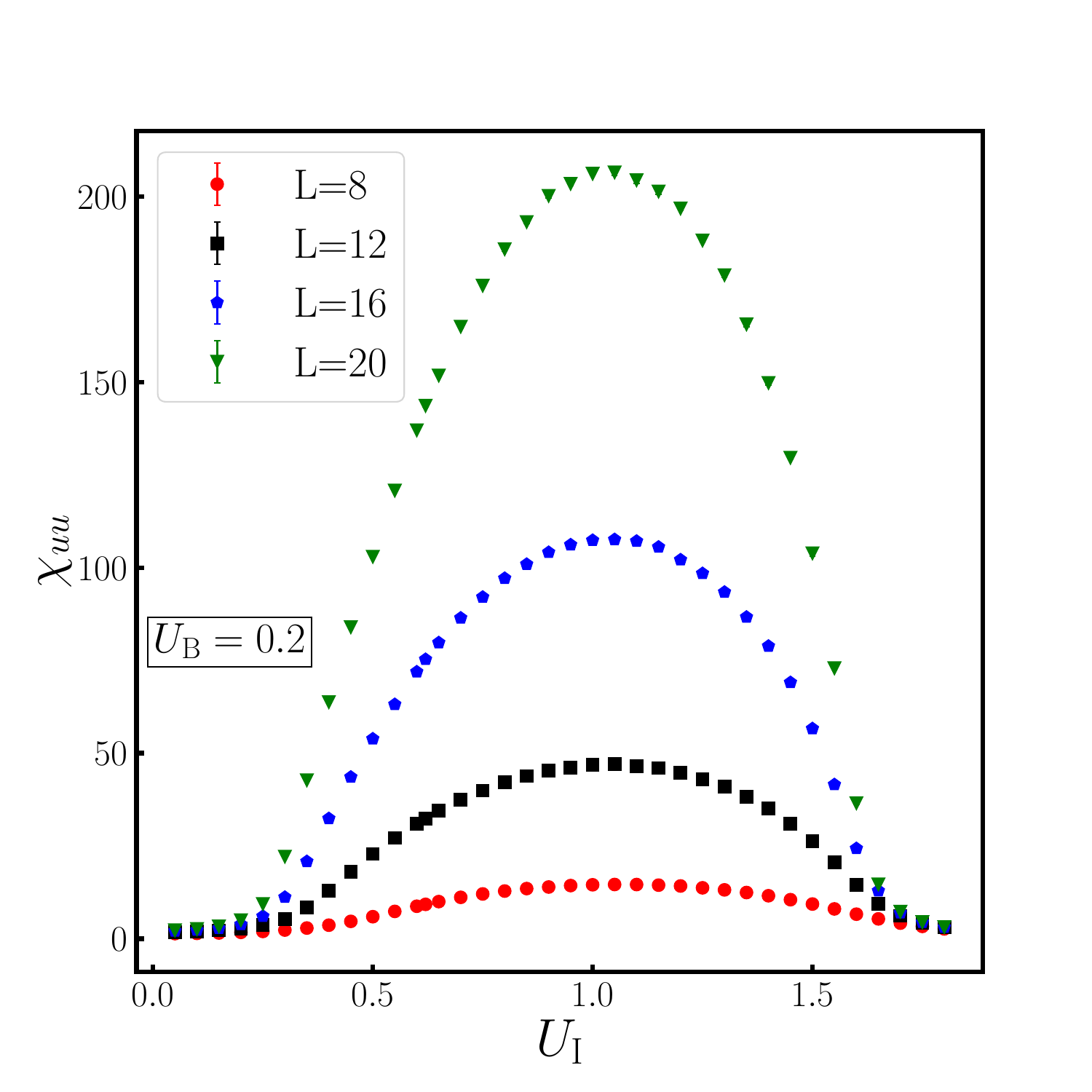}
}
\caption{Plots of $\chi_{ud}$ (top row) and $\chi_{uu}$ (bottom row) as a function of $U_I$ for $U_B = 0.0$ (left), $0.1$(center) and $0.2$ (right).}
\label{fig:phase-chi}
\end{figure*}

\begin{figure}
\centering
\includegraphics[width=0.49\textwidth]{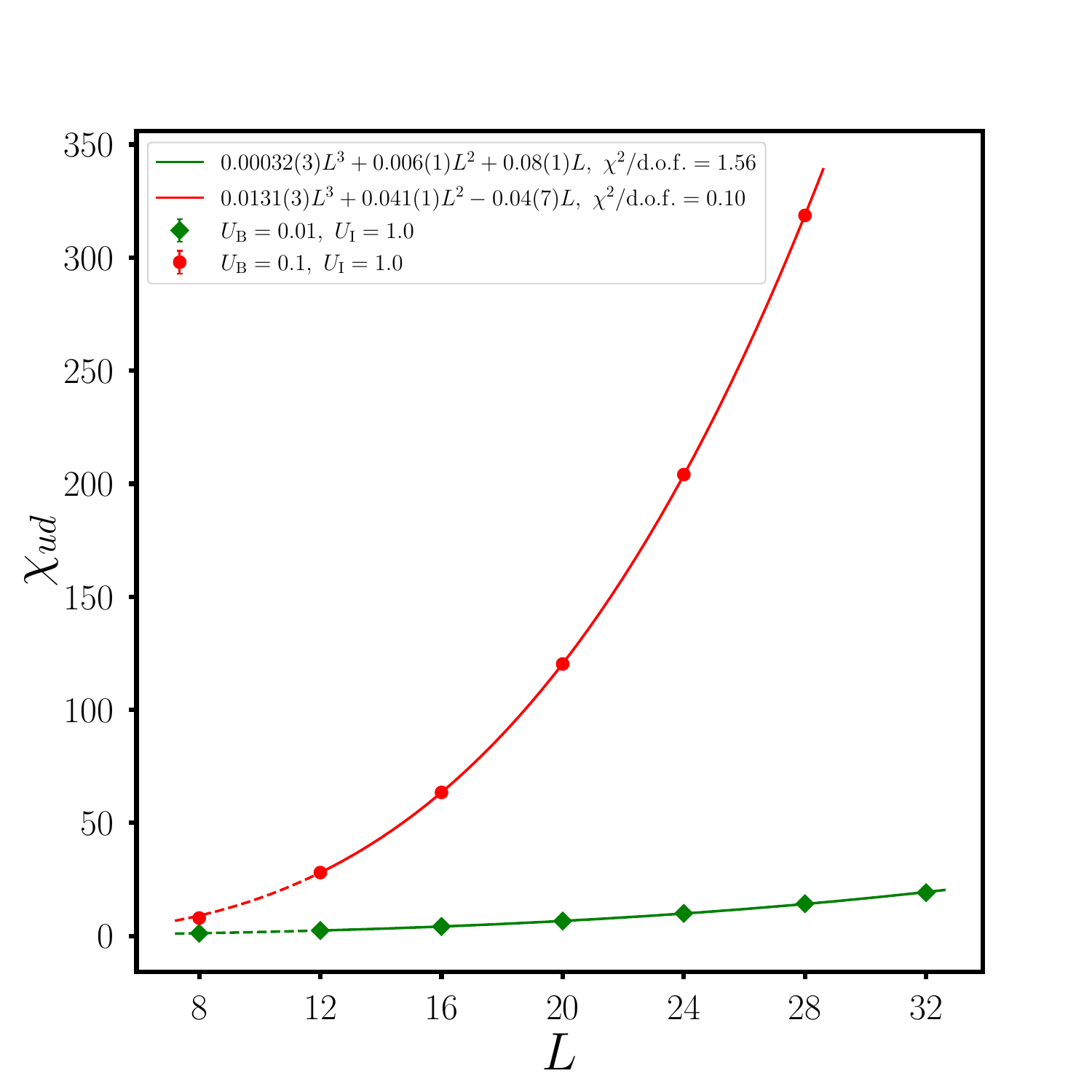}
\caption{Plot of $\chi_{ud}$ as a function of $L$ at $U_I = 1.0$ for $U_B = 0.1$ and $U_B = 0.01$. The solid lines show fits to the form $\chi_{ud} = a L^3 + b L^2 + c L$, as motivated by chiral perturbation theory. The observed scaling is consistent with the existence of an intermediate \ac{SSB} phase for nonzero $U_B$. The solid line shows the range of data used in the fit ($12\leq L \leq 32)$, while the dashed line is the extrapolation of the fit to lower lattice sizes $(8\leq L \leq 12)$.}
\label{fig:chi-ssb}
\end{figure}

\subsection{Phase Diagram}
\label{sec-5b}

One of the main motivations for our study is to understand how a nonzero $U_B$ modifies the direct transition between the \ac{MF} phase at small $U_I$ and the \ac{SMG} phase at large $U_I$ that was found in earlier studies at $U_B=0$ \cite{Ayyar:2014eua,Ayyar:2015lrd}. Based on earlier work \cite{Hasenfratz:1988vc,Lee:1989mi,Hasenfratz:1989jr,Bock:1990tv}, we anticipate the appearance of a conventional massive-fermion phase characterized by \ac{SSB}, as discussed in \cref{sec-1}. A central question is whether this phase emerges only beyond a finite threshold in $U_B$, or whether it appears already for arbitrarily small nonzero $U_B$. In the latter case, the direct \ac{MF}--\ac{SMG} transition at $U_B=0$ would correspond to a multicritical point at which two distinct phase boundaries merge.

Such a multicritical interpretation is further supported by symmetry considerations. At $U_B=0$, the symmetry of the lattice model is enhanced from $\SU(2)\times\SU(2)\times U_\chi(1)$ to $\SU(4)$ (see \cref{sec-2}), which allows the two transitions that are distinct for $U_B \neq 0$ to merge into a single critical point.

To explore the presence of the \ac{SSB} phase, we examine the susceptibilities $\chi_{ud}$ and $\chi_{uu}$, which are expected to scale proportionally to the spacetime volume, $L^3$, when the continuous $U_\chi(1)$ symmetry of the model is spontaneously broken. In \cref{fig:phase-chi}, we show these susceptibilities as functions of $U_I$ for $U_B = 0$, $0.1$, and $0.2$, and for lattice sizes $L = 8, 12, 16$, and $20$. For both small and large values of $U_I$, the susceptibilities saturate as $L$ increases, which is consistent with symmetric phases. In contrast, within an intermediate range of $U_I$, both susceptibilities grow rapidly with system size, signaling spontaneous symmetry breaking.

\begin{figure}[t]
\centering
\includegraphics[width=0.48\textwidth]{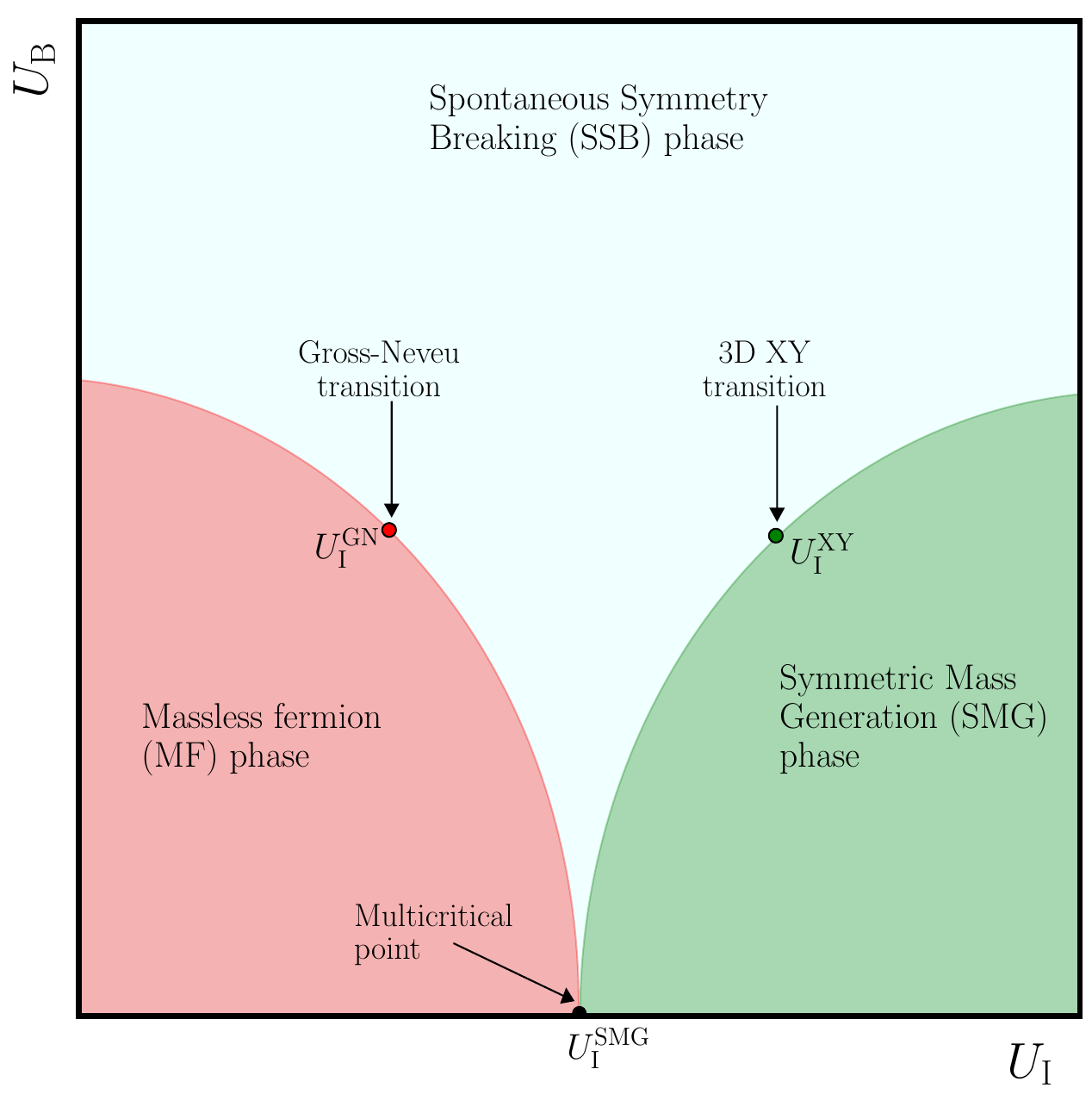}
\caption{Schematic phase diagram of the model defined in \cref{eq:model}. For nonzero values of both $U_B$ and $U_I$, the model possesses a $\SU(2)\times \SU(2)\times U_\chi(1)$ symmetry, which is enhanced to $\SU(4)$ along the $U_B = 0$ axis. Our results suggest that the direct second-order transition on this axis between the \ac{MF} and \ac{SMG} phases corresponds to a multicritical point where the three-dimensional Gross--Neveu and three-dimensional XY critical lines meet. In the \ac{SSB} phase, the $U_\chi(1)$ symmetry is spontaneously broken.}
\label{fig:phase-diag}
\end{figure}

To verify that this growth is consistent with the expected volume scaling, we plot $\chi_{ud}$ as a function of $L$ at $U_I = 1$ for $U_B = 0.1$ and $0.01$ in \cref{fig:chi-ssb}. We find that the data for $L \geq 12$ are well described by the functional form
\begin{align}
\chi_{ud} = a L^3 + b L^2 + c L,
\end{align}
which corresponds to the expected finite-size behavior in the spontaneously broken phase as predicted by chiral perturbation theory \cite{Hasenfratz:1989pk}. These results provide compelling evidence that a symmetry-broken massive-fermion phase emerges for arbitrarily small nonzero $U_B$.

The existence of this phase implies that, for $U_B \neq 0$, the system undergoes two separate continuous phase transitions: a transition between the \ac{MF} and \ac{SSB} phases belonging to the three-dimensional Gross--Neveu-XY universality class, and a second transition between the \ac{SSB} and \ac{SMG} phases in the three-dimensional XY universality class. The single critical point observed at $U_B=0$ can therefore be interpreted as a multicritical point at which these two critical lines merge due to the enhanced $\SU(4)$ symmetry. Based on these observations, our conjectured phase diagram for the lattice model defined in \cref{eq:model} is shown in \cref{fig:phase-diag}. In the following subsections, we analyze each of these transitions in detail.

\begin{table*}
\renewcommand{\arraystretch}{2.0}
\setlength{\tabcolsep}{3.5pt}    
\centering
\begin{tabular}{c c c c c c c c c c cc}
\TopRule
& $f_0$ & $f_1$ & $f_2$ & $f_3$ & $f_4$ & $L$-range & ${U_I}^{\rm GN}$ & $\nu$ & $\eta$  & $\chi^2/d.o.f$ & fit-type\\
\MidRule
$\chi_{uu}$ &  0.19(1) & 0.10(1) & 0.055(4) & 0.032(4) & 0.011(2) & 
\multirow{2}{*}{24-56}
& \multirow{2}{*}{0.610(1)}
 & \multirow{2}{*}{1.04(1)} & 
 \multirow{2}{*}{1.00(2)} & \multirow{2}{*}{1.57} & 
 \multirow{2}{*}{(a)}
 \\
$\chi_{ud}$ & 0.17(1)  & 0.10(1) & 0.055(4) & 0.033(4) & 0.011(4) &  &  &  &  &  & \\
\MidRule
$\chi_{uu}$ &  0.15(1) & 0.08(1) & 0.041(4) & 0.020(3) & 0.005(1) & 
\multirow{2}{*}{24-56}
& \multirow{2}{*}{0.614(1)}
 & \multirow{2}{*}{1.01(1)} & 
 \multirow{2}{*}{0.92(2)} & \multirow{2}{*}{8.2} & 
  \multirow{2}{*}{(b)} \\
$\chi_{ud}$ & 0.15(1)  & 0.08(1) & 0.041(4) & 0.020(3) & 0.005(1) &  &  &  &  & \\
\MidRule
$\chi_{uu}$ &  0.1883(5) & 0.0911(6) & 0.0438(7) & 0.0218(3) & 0.0060(3) & 
\multirow{2}{*}{24-56}
& \multirow{2}{*}{0.6102(1)}
 & \multirow{2}{*}{1.0} & 
 \multirow{2}{*}{1.0} & \multirow{2}{*}{1.7} &
  \multirow{2}{*}{(c)}
 \\
$\chi_{ud}$ & 0.1770(5)  & 0.0901(6) & 0.0433(7) & 0.0226(3) & 0.0060(3) &  &  &  &  & & \\
\MidRule
$\chi_{uu}$ &  0.19(1) & 0.11(1) & 0.06(1) & 0.04(1) & 0.015(4) & 
\multirow{2}{*}{28-56}
& \multirow{2}{*}{0.610(1)}
 & \multirow{2}{*}{1.06(2)} & 
 \multirow{2}{*}{1.00(3)} & \multirow{2}{*}{1.46} &
  \multirow{2}{*}{(a)} \\
$\chi_{ud}$ & 0.18(2)  & 0.11(2) & 0.06(1) & 0.04(1) & 0.014(4) &  &  &  &  & & \\
\MidRule
$\chi_{uu}$ &  0.15(2) & 0.09(1) & 0.05(1) & 0.03(1) & 0.007(4) & 
\multirow{2}{*}{28-56}
& \multirow{2}{*}{0.613(1)}
 & \multirow{2}{*}{1.02(2)} & 
 \multirow{2}{*}{0.94(3)} & \multirow{2}{*}{7.4} &
  \multirow{2}{*}{(b)} \\
$\chi_{ud}$ & 0.15(2)  & 0.09(1) & 0.05(1) & 0.03(1) & 0.007(4) &  &  &  &  & & \\
\MidRule
$\chi_{uu}$ &  0.1891(6) & 0.0906(6) & 0.0429(7) & 0.0230(3) & 0.0067(2) & 
\multirow{2}{*}{28-56}
& \multirow{2}{*}{0.6105(1)}
 & \multirow{2}{*}{1.0} & 
 \multirow{2}{*}{1.0} & \multirow{2}{*}{1.6} &
  \multirow{2}{*}{(c)} \\
$\chi_{ud}$ & 0.1781(5)  & 0.0895(6) & 0.0440(7) & 0.0237(3) & 0.0062(2) &  &  &  &  & & \\
\MidRule
$\chi_{uu}$ &  0.17(2) & 0.10(2) & 0.05(1) & 0.03(1) & 0.009(5) & 
\multirow{2}{*}{32-56}
& \multirow{2}{*}{0.612(2)}
 & \multirow{2}{*}{1.03(3)} & 
 \multirow{2}{*}{0.97(4)} & \multirow{2}{*}{1.36} &
  \multirow{2}{*}{(a)} \\
$\chi_{ud}$ & 0.17(2)  & 0.10(2) & 0.05(1) & 0.03(1) & 0.008(4) &  &  &  &  & & \\
\MidRule
$\chi_{uu}$ &  0.14(2) & 0.07(1) & 0.04(1) & 0.02(1) & 0.003(2) & 
\multirow{2}{*}{32-56}
& \multirow{2}{*}{0.615(2)}
 & \multirow{2}{*}{0.99(3)} & 
 \multirow{2}{*}{0.90(5)} & \multirow{2}{*}{6.2} & 
  \multirow{2}{*}{(b)} \\
$\chi_{ud}$ & 0.14(2)  & 0.07(1) & 0.04(1) & 0.02(1) & 0.003(2) &  &  &  &  & & \\
\MidRule
$\chi_{uu}$ &  0.1902(7) & 0.0920(8) & 0.0439(9) & 0.0224(4) & 0.0063(3) & 
\multirow{2}{*}{32-56}
& \multirow{2}{*}{0.6108(2)}
 & \multirow{2}{*}{1.0} & 
 \multirow{2}{*}{1.0} & \multirow{2}{*}{1.4} & 
  \multirow{2}{*}{(c)} \\
$\chi_{ud}$ & 0.1797(7)  & 0.0913(8) & 0.0459(9) & 0.0226(4) & 0.0054(3) &  &  &  &  & & \\
\MidRule
$\chi_{uu}$ &  0.17(3) & 0.09(2) & 0.05(2) & 0.03(2) & 0.008(7) & 
\multirow{2}{*}{36-56}
& \multirow{2}{*}{0.612(3)}
 & \multirow{2}{*}{1.03(4)} & 
 \multirow{2}{*}{0.96(7)} & \multirow{2}{*}{1.39} &
  \multirow{2}{*}{(a)} \\
$\chi_{ud}$ & 0.16(3)  & 0.09(2) & 0.05(2) & 0.03(2) & 0.007(6) &  &  &  &  & & \\
\MidRule
$\chi_{uu}$ &  0.190(1) & 0.092(1) & 0.044(1) & 0.023(1) & 0.006(0) & 
\multirow{2}{*}{36-56}
& \multirow{2}{*}{0.611(0)}
 & \multirow{2}{*}{1.0} & 
 \multirow{2}{*}{1.0} & \multirow{2}{*}{1.4} &
  \multirow{2}{*}{(c)} \\
$\chi_{ud}$ & 0.181(1)  & 0.091(1) & 0.045(1) & 0.023(1) & 0.006(0) &  &  &  &  & & \\
\BotRule
\end{tabular} 
\caption{Combined fit results for $\chi_{uu}$ and $\chi_{ud}$ in the critical region for various lattice ranges. We perform three types of fits: (a) all parameters ${U_I}^{\rm GN}$, $\nu$, $\eta$, and polynomial coefficients  treated as independent; (b) all parameters constrained to be identical for both observables; (c) fits with the constraint $\nu=\eta=1$. In all cases we assume a quartic universal scaling function of the form given in \cref{eq:univfn-GN}.}
\label{tab:GN-chifits}
\end{table*}

\begin{figure}[t]
\centering
\includegraphics[width=0.48\textwidth]{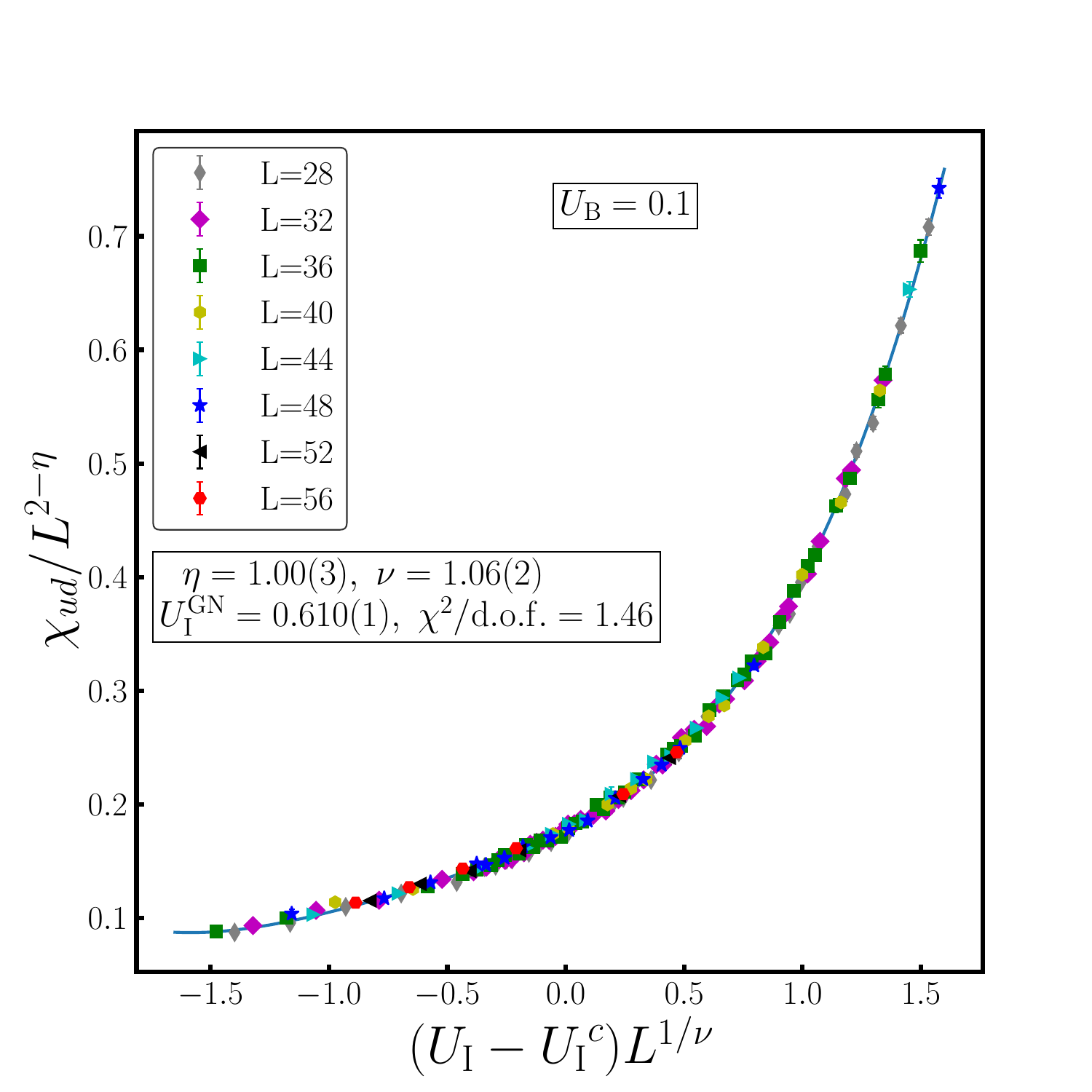}
\caption{Plot of the critical finite-size scaling function defined in \cref{eq:crit-scaling} at the Gross--Neveu transition for $U_B = 0.1$. The solid line represents a fourth-order polynomial fit.}
\label{fig:GNFSS}
\end{figure}

\subsection{Gross--Neveu Transition}
\label{sec-5c}

If the phase transition from the \ac{MF} phase to the \ac{SSB} phase in our lattice model is continuous, the long-distance physics near criticality should be governed by the three-dimensional Gross--Neveu universality class. This universality class has been studied extensively using a wide range of analytical and numerical approaches \cite{Rosenstein:1990nm,Hands:1992be,Rosenstein:1993zf,Hofling:2002hj,Herbut:2006cs}. Determining its universal properties, in particular the critical exponents, remains an active area of research \cite{Gracey2025}. These universal properties depend on several factors, including the number of fermion flavors, the pattern of symmetry breaking, and lattice-specific features such as fermion doubling. As a result, when staggered fermions are used, the mapping between lattice symmetries and continuum symmetries must be handled with care.

A single flavor of staggered fermions on a three-dimensional cubic lattice is expected to describe two flavors of four-component Dirac fermions in the continuum limit. Since the lattice model defined in \cref{eq:model} contains two flavors of staggered fermions, the \ac{MF} phase is therefore expected to correspond to four flavors of four-component Dirac fermions at long distances. Moreover, because the \ac{SSB} phase breaks only the $U_\chi(1)$ subgroup of the full $\SU(2)\times \SU(2)\times U_\chi(1)$ symmetry of the lattice model, the transition between the \ac{MF} and \ac{SSB} phases is expected to be in the Gross--Neveu--XY universality class with four Dirac fermion flavors.

A characteristic prediction of this universality class is that the fermion bilinear susceptibilities $\chi_{ud}$ and $\chi_{uu}$ obey the finite-size scaling relation
\begin{align}
\chi\, / L^{2-\eta} = f\!\left((U_I - U_I^c)\, L^{1/\nu}\right),
\label{eq:crit-scaling}
\end{align}
in the vicinity of the critical point. The universal scaling function $f(x)$, where $x = (U_I - U_I^c)\, L^{1/\nu}$, may differ for the two susceptibilities, although in our fits they are almost identical. At the Gross--Neveu transition we denote $U_I^c = U_I^{\rm GN}$ and determine this critical coupling, along with the critical exponents $\eta$ and $\nu$, by performing a combined fit of both $\chi_{ud}$ and $\chi_{uu}$ data to the scaling form in \cref{eq:crit-scaling}. For each susceptibility, we approximate the scaling function by a polynomial expansion,
\begin{align}
f(x) = f_0 + f_1 x + f_2 x^2 + f_3 x^3 + f_4 x^4,
\label{eq:univfn-GN}
\end{align}
and study the stability of the fits under variations of the fitting range in lattice size. The resulting fit parameters are summarized in \cref{tab:GN-chifits}. The data for $\chi_{ud}$ together with a representative scaling fit are shown in \cref{fig:GNFSS}.

\begin{table}[h]
\centering
\renewcommand{\arraystretch}{1.4}
\setlength{\tabcolsep}{4pt}
\begin{tabular}{l|l|c}
\TopRule
$\nu$ & $\eta$ & Citation\\
\MidRule
$1.06(2)$ & $1.00(3)$ & This work 
\\
$1.13$ & $0.93(09)$ & MC study (bilayer graphene) \cite{Nature2025-Huang} \\
$1.06(9)$ & $0.99(5)$ &
$\varepsilon$-expansion (4-loop)
\cite{PhysRevD.96.096010} \\
$1.092(6)$ & $0.926(13)$ & $\varepsilon$-expansion (interpolation) \cite{PhysRevB.111.205129} \\
\BotRule 
\end{tabular}
\caption{Critical exponents reported in various studies for the Gross--Neveu phase transition in three Euclidean dimensions with four flavors of four-component Dirac fermions. At this transition, massless fermions acquire a mass through the formation of a fermion--bilinear condensate that spontaneously breaks a $U(1)$ symmetry of the theory. The results from earlier studies are summarized in Table~I of Ref.~\cite{PhysRevB.111.205129}.
\label{tab:GN-exp} }
\end{table}

\begin{figure*}[t]
\centering
\includegraphics[width=0.48\textwidth]{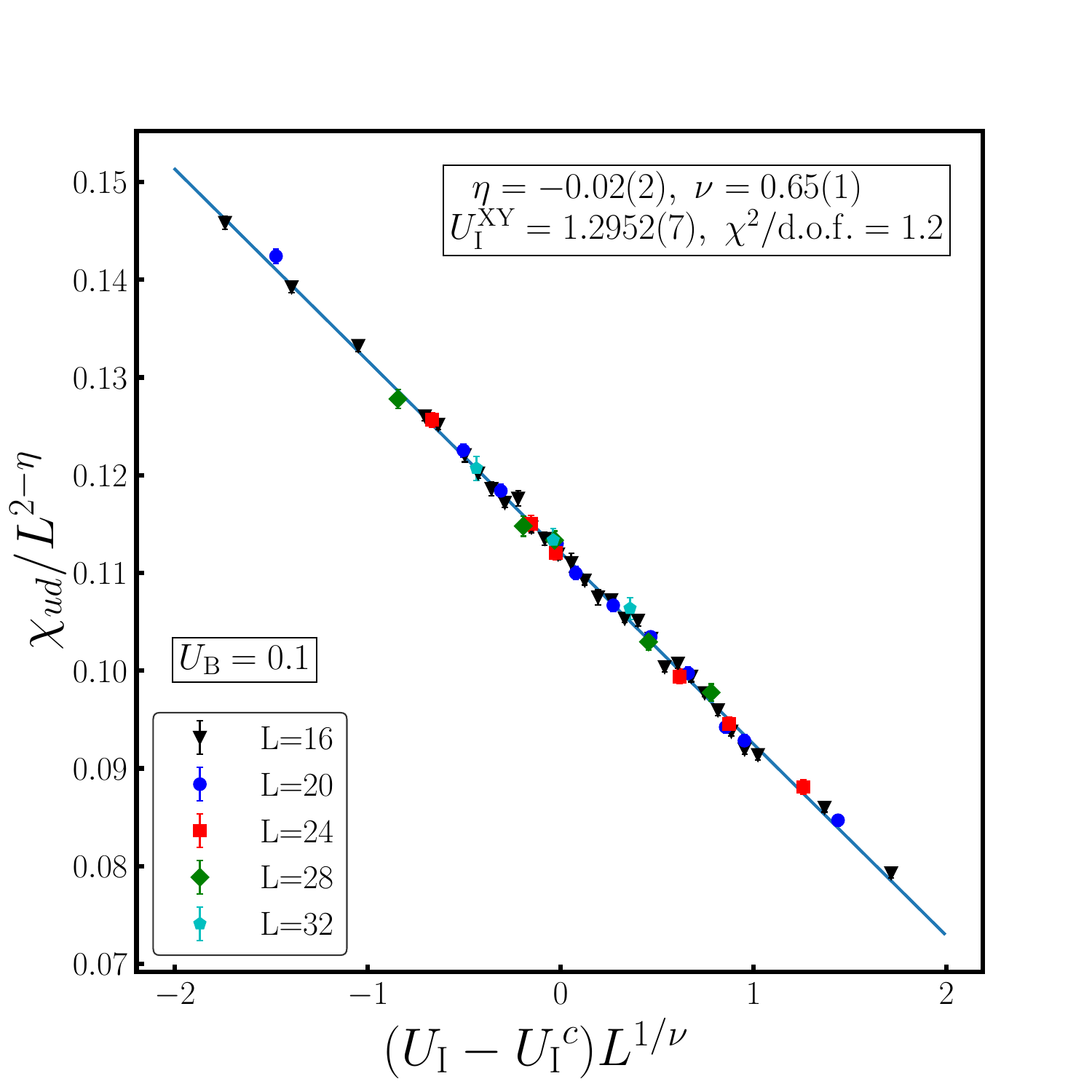}
\includegraphics[width=0.48\textwidth]{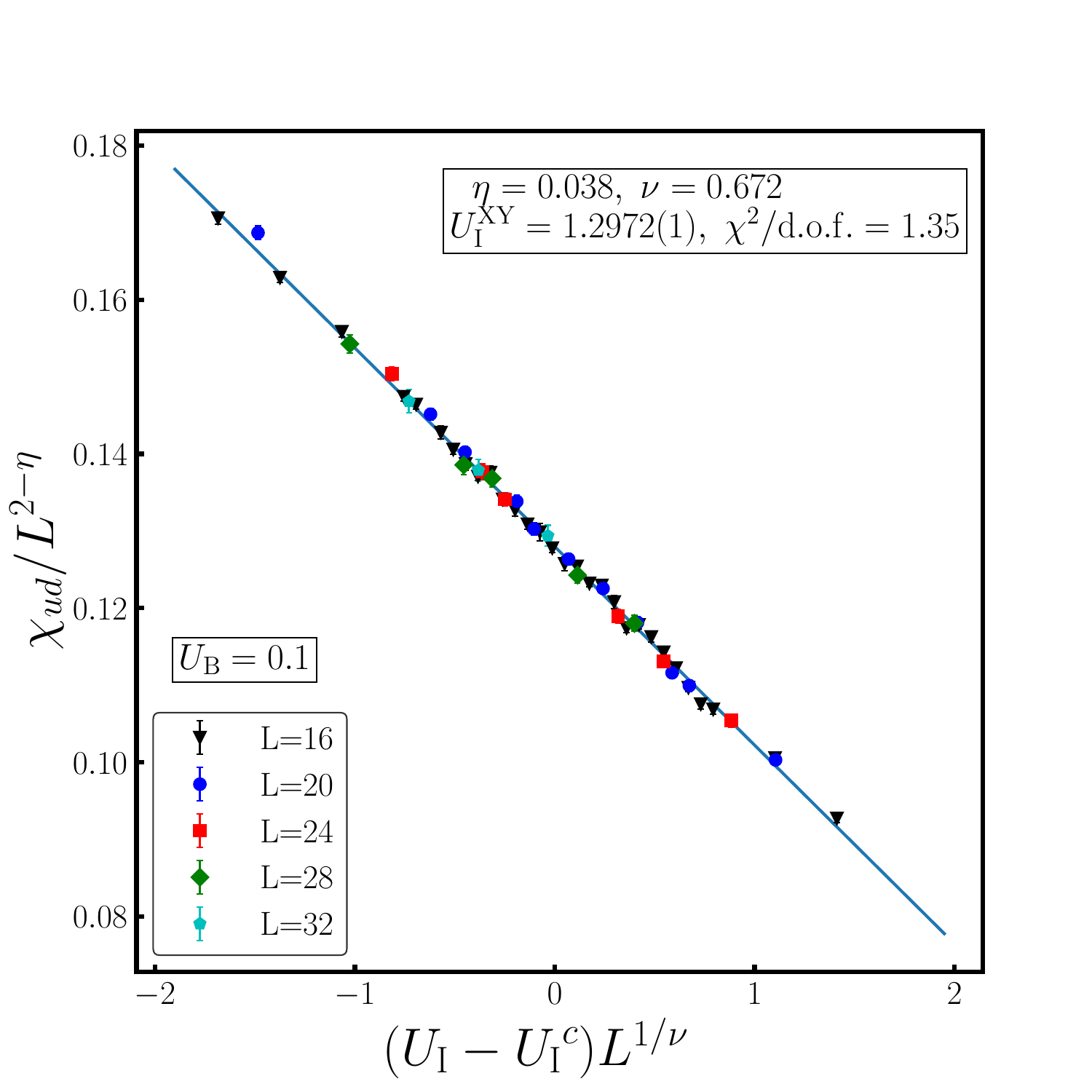}
\caption{Finite-size scaling analysis of the critical scaling function defined in \cref{eq:univfn-XY} at the XY transition for $U_B = 0.1$. Two different scaling analyses are shown. In the left panel, the critical coupling $U_I^c$ and the critical exponents are treated as fit parameters, and the data are fitted using a linear approximation \cref{eq:univfn-XY}. In the right panel, the critical exponents are fixed to their three-dimensional XY values \cite{PhysRevB.63.214503}, and the data are fitted with respect to $U_I^c$ and the coefficients of the linear scaling function. 
}
\label{fig:XYFSS}
\end{figure*}

\begin{table*}
\renewcommand{\arraystretch}{1.9}
\setlength{\tabcolsep}{2.8pt}   
\centering
\begin{tabular}{c c c c c c c c c}
\TopRule
& $f_0$ & $f_1$ & $L$-range & ${U_I}^{\rm XY}$ & $\nu$ & $\eta$  & $\chi^2/d.o.f$ & fit-type\\
\MidRule
$\chi_{uu}$ & 0.113(2) & -0.0179(7) & \multirow{2}{*}{12-32}
& \multirow{2}{*}{1.2953(3)}
 & \multirow{2}{*}{0.644(6)} & 
 \multirow{2}{*}{-0.02(1)} & \multirow{2}{*}{1.0} &
  \multirow{2}{*}{(a)} \\
$\chi_{ud}$ & 0.112(2) & -0.0178(7) & & & & & \\
\MidRule
$\chi_{uu}$ & 0.112(2) & -0.018(1) & \multirow{2}{*}{12-32}
& \multirow{2}{*}{1.2953(3)}
 & \multirow{2}{*}{0.644(6)} & \multirow{2}{*}{-0.02(1)} 
 & \multirow{2}{*}{3.8} & 
 \multirow{2}{*}{(b)} \\
$\chi_{ud}$ & 0.112(2) & -0.018(1) & & & & & \\
\MidRule
$\chi_{uu}$ & 0.1270(1) & -0.0241(1) & \multirow{2}{*}{12-32}
& \multirow{2}{*}{1.2979(1)}
 & \multirow{2}{*}{0.672} & \multirow{2}{*}{0.038} 
 & \multirow{2}{*}{1.46} &
  \multirow{2}{*}{(c)} \\
$\chi_{ud}$ & 0.1270(1) & -0.0241(1) & & & & & & \\
\MidRule
$\chi_{uu}$ & 0.113(5) & -0.020(2) & \multirow{2}{*}{16-32}
& \multirow{2}{*}{1.2952(7)}
 & \multirow{2}{*}{0.65(1)} & 
 \multirow{2}{*}{-0.02(2)} & \multirow{2}{*}{1.2} & 
  \multirow{2}{*}{(a)} \\
$\chi_{ud}$ & 0.112(5) & -0.020(2) & & & & & \\
\MidRule
$\chi_{uu}$ & 0.112(5) & -0.020(2) & \multirow{2}{*}{16-32}
& \multirow{2}{*}{1.2952(7)}
 & \multirow{2}{*}{0.65(1)} & \multirow{2}{*}{-0.02(2)} 
 & \multirow{2}{*}{1.4} & 
  \multirow{2}{*}{(b)} \\
$\chi_{ud}$ & 0.112(5) & -0.020(2) & & & & & \\
\MidRule
$\chi_{uu}$ & 0.1287(2) & -0.0257(1) & \multirow{2}{*}{16-32}
& \multirow{2}{*}{1.2972(1)}
 & \multirow{2}{*}{0.672} & \multirow{2}{*}{0.038} 
 & \multirow{2}{*}{1.35} & 
  \multirow{2}{*}{(c)} \\
$\chi_{ud}$ & 0.1280(2) & -0.0258(1) & & & & & \\
\BotRule
\end{tabular} 
\caption{Combined fit results for $\chi_{uu}$ and $\chi_{ud}$ in the critical region for various lattice ranges. We perform three types of fit: (a) all parameters ${U_I}^{\rm XY}$, $\nu$, $\eta$, and polynomial coefficients treated as independent; (b) all parameters constrained to be identical for both observables; (c) fits performed with the three-dimensional XY critical exponents fixed to $\nu = 0.672$ and $\eta = 0.038$. In all cases we assume a linear universal scaling function of the form given in \cref{eq:univfn-XY}.}
\label{tab:XY-chifits}
\end{table*}

\begin{figure*}[t]
\centering
\includegraphics[width=0.50\textwidth]{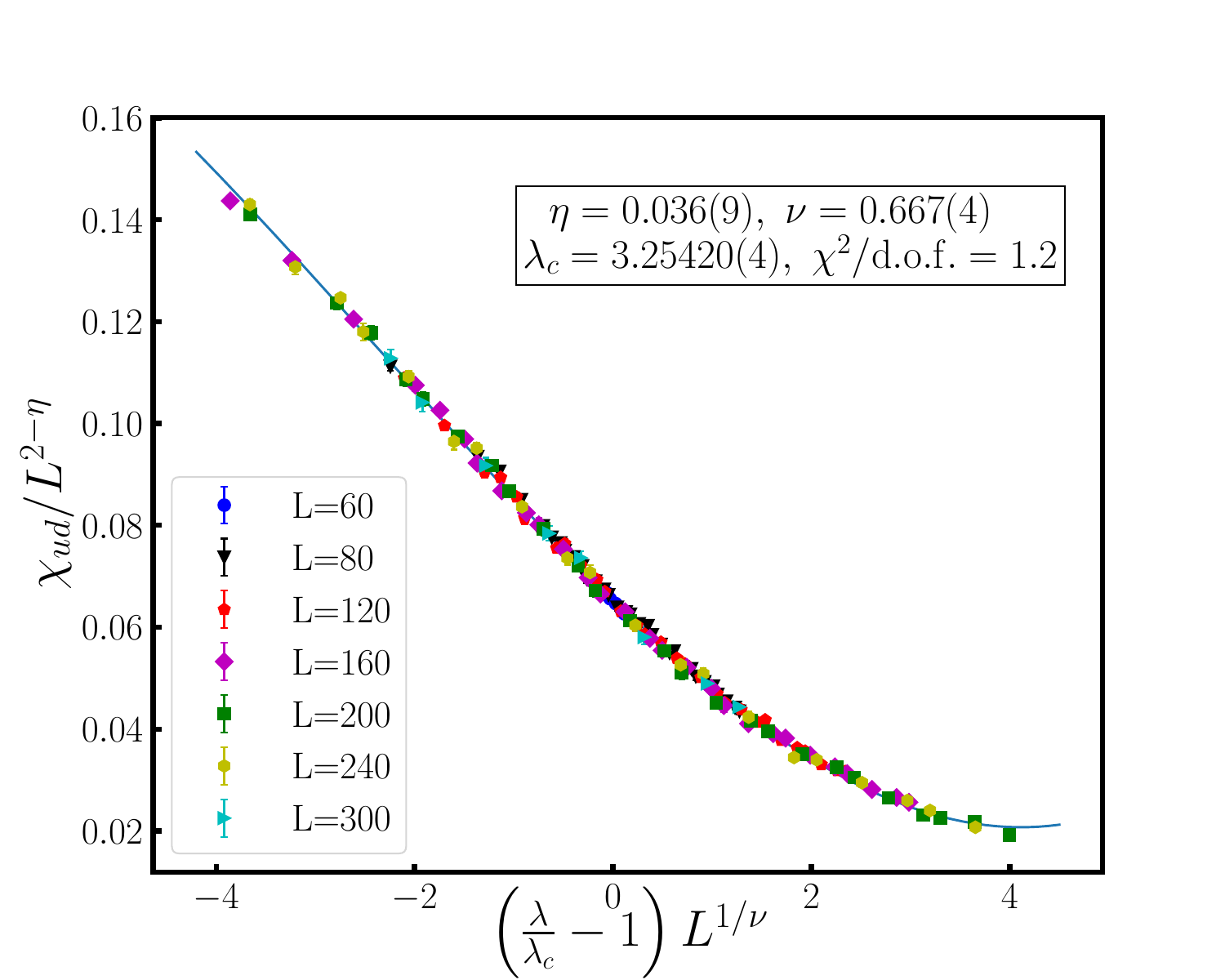}
\caption{Finite-size scaling analysis of the 3D XY transition in the regime of large 
$U_B$ and $U_I$, with $\lambda = U_I/U_B$. The scaling variable is $x=(\lambda/{\lambda}_c - 1)L^{1/\nu}$. The solid lines show fits to the scaling function defined in \cref{eq:crit-scaling}, including terms up to order $x^3$. The extracted critical exponents are in good agreement with those of the three-dimensional XY universality class. The fitted coefficients are $f_0 = 0.006(2),\ f_1 = -0.019(1),\ f_2 = 0.0012(1),\ \text{and}\ f_3 = 0.00017(2)$.}
\label{fig:3DXYFSS}
\end{figure*}

From this analysis, we obtain the estimates $U_I^{\rm GN} = 0.610(1)$, $\nu = 1.06(2)$, and $\eta = 1.00(3)$. These values are in reasonable agreement with critical exponents reported in previous analytical and numerical studies of the $U(1)$ Gross--Neveu universality class with four Dirac fermion flavors, as summarized in \cref{tab:GN-exp}.

\subsection{3D-XY Transition}
\label{sec-5d}

If the phase transition from the \ac{SSB} phase to the \ac{SMG} phase is continuous, a natural expectation is that it belongs to the three-dimensional XY universality class, which has been extensively studied \cite{PhysRevB.63.214503}. This expectation is based on the observation that fermionic excitations are massive throughout the \ac{SSB} phase, so that the long-distance critical behavior should be governed by bosonic modes associated with the spontaneous breaking of the $U_\chi(1)$ symmetry.

Nevertheless, it is not a priori guaranteed that all effects of the underlying fermionic degrees of freedom are irrelevant at the transition. In particular, one may ask whether remnants of the microscopic fermionic structure, such as topological effects, could modify the universality class. To address this possibility, we analyze the second transition using the fermion-bag Monte Carlo approach. Near this transition, the fermion bilinear susceptibility is again expected to satisfy the finite-size scaling form in \cref{eq:crit-scaling}, with the critical coupling identified as $U_I^c = U_I^{\rm XY}$.

From a computational perspective, this transition is more challenging to study than the Gross--Neveu transition, since it occurs within a microscopic fermionic model rather than a purely bosonic one. In certain limits, the problem can be simplified by taking both $U_I$ and $U_B$ to be large while keeping the ratio $\lambda = U_I/U_B$ finite. Such a limit was recently studied in two spacetime dimensions, where a Berezinskii--Kosterlitz--Thouless transition was observed as $\lambda \to 0$ \cite{PhysRevLett.132.041601}. A similar simplification can in principle be considered in three dimensions by taking both couplings large at fixed $\lambda$, in which case the system falls into the three-dimensional XY universality class, as demonstrated by the results shown in \cref{fig:3DXYFSS}. In the limit of large $U_I$ and $U_B$, 
the fermions completely decouple and the lattice models become purely bosonic. For this reason, simulations on much larger lattices are computationally feasible. We therefore extend the analysis up to $L=300$ to further suppress finite-size corrections and improve the stability of the data-collapse analysis. However, if we do not explore that limit and instead focus on the more generic parameter regime, our Monte Carlo study of this transition is restricted to lattice sizes up to $L = 32$ as shown in \cref{fig:XYFSS}.

Given these limitations, we focus on a relatively narrow critical region where the universal scaling function $f(x)$ can be approximated by its leading linear form,
\begin{align}
f(x) = f_0 + f_1 x,
\label{eq:univfn-XY}
\end{align}
which is sufficient for the lattice sizes accessible in our simulations.

We perform two complementary analyses: one in which the critical exponents are treated as free fitting parameters, and another in which they are fixed to their three-dimensional XY values. A summary of the resulting fits, together with additional fitting strategies, is provided in \cref{tab:XY-chifits}. The corresponding scaling plots for the two analyses are shown in \cref{fig:XYFSS}. Since the exponents obtained from fits with free parameters are close to the known three-dimensional XY values within uncertainties, a conservative interpretation of our results is that the \ac{SSB}--\ac{SMG} transition lies in the standard three-dimensional XY universality class.

In summary, our finite-size scaling analysis of the \ac{SSB}--\ac{SMG} transition yields results consistent with a continuous transition in the three-dimensional XY universality class. While the microscopic fermionic nature of the model and the accessible lattice sizes limit the precision of the analysis, the observed scaling behavior of the fermion bilinear susceptibilities and the proximity of the fitted critical exponents to their known three-dimensional XY values support this interpretation. Nevertheless, as $U_B$ is reduced and fermionic degrees of freedom become increasingly important, additional effects may influence the critical behavior. More detailed studies---particularly in the regime of small $U_B$---would therefore be valuable to further elucidate the multicritical behavior that emerges at $U_B=0$.

\section{Conclusions}
\label{sec-6}

In this work, we have investigated the phase structure and critical behavior of a three-dimensional lattice model of two flavors of massless staggered fermions interacting through two independent four-fermion couplings, $U_I$ and $U_B$, using the fermion-bag Monte Carlo approach. At $U_B = 0$, this model is known to exhibit a direct continuous transition between a massless fermion (\ac{MF}) phase and a symmetric massive fermion (\ac{SMG}) phase, where fermions acquire mass without the formation of a fermion bilinear condensate. Our primary goal has been to understand how this unconventional transition is modified when a second interaction channel, $U_B$, is introduced.

Our results show that an intermediate phase with spontaneous symmetry breaking (\ac{SSB}) emerges as soon as $U_B$ becomes nonzero. This phase is characterized by a nonvanishing fermion bilinear condensate that breaks the $U_\chi(1)$ symmetry of the lattice model. The appearance of this phase splits the single transition observed at $U_B = 0$ into two distinct continuous transitions: a transition between the \ac{MF} and \ac{SSB} phases, and a second transition between the \ac{SSB} and \ac{SMG} phases.

We have analyzed the critical behavior of the \ac{MF}--\ac{SSB} transition using finite-size scaling of fermion bilinear susceptibilities. Our results are consistent with this transition belonging to the three-dimensional $U(1)$ Gross--Neveu universality class with four flavors of four-component Dirac fermions, in agreement with expectations based on symmetry considerations and earlier studies. For the \ac{SSB}--\ac{SMG} transition, we performed a complementary finite-size scaling analysis and found behavior consistent with the three-dimensional XY universality class. While the microscopic fermionic nature of the model and the accessible lattice sizes limit the precision of this analysis, the extracted critical exponents are close to their known three-dimensional XY values.

Taken together, these results suggest that the direct \ac{MF}--\ac{SMG} transition observed at $U_B = 0$ can be interpreted as a multicritical point at which the Gross--Neveu and XY critical lines merge. This interpretation is further supported by the enhancement of the lattice symmetry from $\SU(2)\times \SU(2)\times U_\chi(1)$ to $\SU(4)$ along the $U_B = 0$ axis. Our findings thus provide a unified picture of how conventional symmetry-breaking transitions and unconventional symmetric mass generation phenomena are connected within a single lattice framework.

Several interesting directions remain for future work. In particular, more detailed studies in the regime of small $U_B$, where fermionic degrees of freedom play an increasingly important role, would provide additional insight into the nature of the multicritical point. Extending the simulations to larger lattice sizes, exploring alternative parameter regimes, or complementary analytical approaches may further shed light on the interplay between enhanced symmetry, fermionic dynamics, and critical behavior in this class of models.

\section*{Acknowledgments}

We are grateful to Subhro Bhattacharjee, Simon Catterall, Asit De, Simon Hands, and Cenke Xu for insightful discussions. S.C. is supported in part by the U.S. Department of Energy, Office of Science, Nuclear Physics program under Award No.~DE-FG02-05ER41368. This research was supported in part by the National Science Foundation under Grant No. NSF PHY-1748958. D.B. would like to thank STFC (UK) consolidated Grant No. ST/X000583/1 and continued support from the Alexander von Humboldt Foundation (Germany) in the context of the
research fellowship for experienced researcher. M.K.M is grateful for the hospitality of Perimeter Institute where part of this work was carried out. Research
at Perimeter Institute is supported in part by the Government of Canada through the Department of Innovation, Science and Economic Development and by the Province of Ontario through the Ministry of Colleges and Universities. This research was also supported in part by the Simons Foundation through the Simons Foundation Emmy Noether Fellows Program at Perimeter Institute. S.M. would like to thank the Institute for Theoretical Physics at ETHZ for hospitality during his visit, where part of this work was carried out. S.M. would like to acknowledge support from the National Natural Science Foundation of China under Grants No.~12325508, No.~12293064, and No.~12293060. D.B., S.M., and M.K.M. also acknowledge the use of computing clusters at SINP, Kolkata, and access to the Piz Daint and Eiger supercomputers at the Swiss National Supercomputing Centre, Switzerland, under ETHZ’s allocation with project IDs \texttt{c21} and \texttt{eth8}. We acknowledge the use of AI assistance, specifically ChatGPT~\hbox{\cite{openai2025chatgpt}}, in refining the language and clarity of this manuscript.

\bibliographystyle{apsrev4-2} 
\showtitleinbib
\bibliography{fixapsbib,refs,smg,GNref}


\appendix
\subsection{Comparison with Exact Results}
\label{app-1}

\begin{table*}[t]
\centering
\renewcommand{\arraystretch}{1.4}
\setlength{\tabcolsep}{4pt}
\begin{tabular}{l|cc|cc|cc r@{.}l c}
\TopRule
$U_I$ & \multicolumn{2}{c}{$\rho_I$} & \multicolumn{2}{c}{$\chi_{uu}$}  & \multicolumn{2}{c}{$\chi_{ud}$} \\
\MidRule
&  Exact & Worm &  Exact & Worm  & Exact & Worm\\
\MidRule
0.05 & 0.0001 & 0.0001(0) & 0.1764 & 0.1768(2) & 0.0021 & 0.0021(0)\\
0.1  & 0.0004 & 0.0004(0) & 0.1765 & 0.1761(2) & 0.0043 & 0.0043(0) \\
0.5  & 0.0098 & 0.0099(0) & 0.1777 & 0.1777(3) & 0.0214 & 0.0213(1) \\
0.6  & 0.0143 & 0.0142(0) & 0.1782 & 0.1783(2) & 0.0256 & 0.0256(1) \\
1.0  & 0.0416 & 0.0418(1) & 0.1809 & 0.1808(2) & 0.0420 & 0.0420(1) \\
\BotRule
\end{tabular}
\caption{ Comparison between exact and Monte Carlo results on a $L=2$ lattice when $U_B=0.1$.}
 \label{Tab:I}
\end{table*}

In this appendix we validate our algorithm by comparing results obtained from Monte Carlo simulations with exact results on an $L=2$ lattice. In addition to the two independent fermion bilinear susceptibilities defined in \cref{eq:obs-chiud} and \cref{eq:obs-chiuu}, we also introduce two additional observables for the comparison: the dimer density,
\begin{align}
\rho_D = \frac{2}{L^3} \sum_{\langle ij\rangle} \langle \ubar_i u_i \, \ubar_j u_j \rangle,
\label{eq:obs-rhoD}
\end{align}
and the instanton density,
\begin{align}
\rho_I = \frac{1}{L^3} \sum_i \langle \ubar_i u_i \, \dbar_i d_i \rangle.
\label{eq:obs-rhoI}
\end{align}
These densities are normalized such that they take values between $0$ and $1$. These densities also help us monitor when our Monte Carlo algorithm has roughly thermalized.

The exact results for the partition function $Z$ (see \cref{eq:pf-def}), the two susceptibilities $\chi_{ud}$ and $\chi_{uu}$, and the two densities $\rho_D$ and $\rho_I$ are given by the following expressions:
\begin{widetext}
\begin{align}
    Z &= 6561 + 34992\;U_B + 97200\;U_B^2 + 181440\;U_B^3 + 245088\;U_B^4 + 241920\;U_B^5 + 172800\;U_B^6 + 82944\;U_B^7 + 20736\;U_B^8 \nonumber \\
     & + U_I^2\;( 972 + 5616\;U_B + 16464\;U_B^2 +  29952\;U_B^3 + 34368\;U_B^4 
     + 23808\;U_B^5 + 7936\;U_B^6) \notag \\
     & + U_I^4\;(126 + 624\;U_B + 1488\;U_B^2  + 1728\;U_B^3 + 864\;U_B^4) + U_I^6\;(12 + 48\;U_B + 48\;U_B^2) + U_I^8, \\
    \rho_I & = \frac{1}{8Z} \big( U_I^2\;(1944 + 11232\;U_B + 32928\;U_B^2 +  59904\;U_B^3 + 68736\;U_B^4 + 47616\;U_B^5 + 15872\;U_B^6) \nonumber \\
    & \qquad\qquad + U_I^4\; (504 + 2496\;U_B 
    + 5952\;U_B^2 + 6912\;U_B^3 + 3456\;U_B^4) + U_I^6\;(72 + 288\;U_B + 288\;U_B^2)
    + 8\;U_I^8 \big),  \\
    \rho_D &= \frac{1}{8Z} \big( U_B\;(34992 + 5616\;U_I^2 + 624\;U_I^4 + 48\;U_I^6) + U_B^2\;( 194400 + 32928\;U_I^2 + 2976\;U_I^4 + 96\;U_I^6) \notag \\
    & \qquad\qquad + U_B^3\;( 544320 + 89856\;U_I^2 + 5184\;U_I^4) + U_B^4\;( 980352 + 137472\;U_I^2 + 3456\;U_I^4) \notag \\
    & \qquad\qquad + U_B^5\;(1209600 + 119040\;U_I^2) 
    + U_B^6\;(1036800 + 47616\;U_I^2) + 580608\;U_B^7 + 165888\;U_B^8 \big), 
    \\
    \chi_{uu} &= \frac{1}{2Z} \big(  (12672\;U_B^7 + 44352\;U_B^6 + 78432\;U_B^5 + 90000\;U_B^4 + 71496\;U_B^3 + 38556\;U_B^2 + 13122\;U_B + 2187) \nonumber\\
    & \qquad\qquad +  U_I^2\;(3744 U_B^5 + 9360\;U_B^4 + 10704\;U_B^3 + 6888\;U_B^2 + 2466\;U_B + 405) \notag \\
    & \qquad\qquad + U_I^4\;(264\;U_B^3 + 396\;U_B^2 + 222\;U_B + 45) + U_I^6\;(6\;U_B + 3 ) \big), \\
    \chi_{ud} &= \frac{1}{2Z} \big(  U_I\;(5760\; U_B^6 + 17280\;U_B^5 + 24672\;U_B^4 + 21024\;U_B^3 + 11112\;U_B^2 + 3456\;U_B + 486) \notag \\ 
    & \qquad \qquad + U_I^3\;(768\;U_B^4 + 1536\;U_B^3 + 1248\;U_B^2 + 480\;U_B + 72) + U_I^5\;(24\;U_B^2 + 24\;U_B + 6) \big).
\end{align}
\end{widetext}
In \cref{Tab:I} we compare the exact values for the above four observables with our Monte Carlo results at $U_B=0.1$ at five different values of $U_I$.

\end{document}